\documentstyle[12pt,aaspp4]{article}
\input psfig.sty

\def \eg           {{e.g.}}
\def \etal         {{et~al. }}
\def \h2         {\hbox{H$_2$}}
\def \ie           {{i.e.}}
\def \IRAS         {\hbox{{\it IRAS\ }}}
\def \kms          {\hbox{km$\,$s$^{-1}$}}
\def\approxlt{\lower.2em\hbox{$\buildrel < \over \sim$}}
\def\approxgt{\lower.2em\hbox{$\buildrel > \over \sim$}}
\def \lco          {\hbox{$L_{\rm CO}$}}
\def \lhcn          {\hbox{$L_{\rm HCN}$}}
\def \ll           {\hbox{K~\kms~pc$^2$}}
\def \lir          {\hbox{$L_{\rm IR}$}}
\def \ls           {\hbox{L$_{\odot}$}}

\def \ms           {\hbox{M$_{\odot}$}}           
\def \nh2          {\hbox{$<$$n_{\rm H_2}$$>$}}
\def \nhtwo        {\hbox{$<$$n({\rm H_2})$$>$}}

\def \date         {\ifcase\month \message{zero} \or
                    January \or February \or March \or April \or May \or June 
                    \or July \or 
                    August \or September \or October \or November \or 
                    December \fi
                    \space\number\day, \number\year}

\begin{document}

\title{HCN Survey of Normal Spiral, IR-luminous and Ultraluminous Galaxies}
\author{Yu Gao$^{1, 2, 3}$ \  and \ Philip M. Solomon$^4$}

\altaffiltext{1}{Purple Mountain Observatory, Chinese Academy of Sciences, 
2 West Beijing Road, Nanjing 210008, P.R. China}
\altaffiltext{2}{Department of Astronomy, University of Massachusetts, \\
LGRT-B 619E, 710 North Pleasant Street, Amherst, MA 01003}
\altaffiltext{3}{Infrared Processing and Analysis Center, 
California Institute of Technology, MS 100-22, Pasadena, CA 91125}
\altaffiltext{4}{Department of Physics \& Astronomy, SUNY at Stony Brook, 
Stony Brook, NY 11794}
\authoremail{gao@astro.umass.edu, psolomon@astro.sunysb.edu}

\begin{abstract}
We report systematic HCN J=1-0 (and CO) observations 
of a sample of 53 infrared (IR) and/or 
CO-bright and/or luminous galaxies, including seven 
ultraluminous infrared galaxies, nearly 20 luminous infrared 
galaxies, and more than a dozen of the nearest
normal spiral galaxies.  This is the largest and most 
sensitive HCN survey of galaxies to date.
All galaxies observed so far follow the tight correlation
between the IR luminosity $L_{\rm IR}$ and the HCN luminosity 
$L_{\rm HCN}$ initially proposed by Solomon, Downes, \& Radford,
which is detailed in a companion paper.
We also address here the issue of HCN excitation.
There is no particularly strong correlation between $L_{\rm HCN}$ 
and the 12$\mu$m luminosity; in fact, of all the 
four \IRAS bands, the 12$\mu$m luminosity has the weakest correlation
with the HCN luminosity.
There is also no evidence of stronger HCN emission or a higher 
ratio of HCN and CO luminosities $L_{\rm HCN}/L_{\rm CO}$
for galaxies with excess 12$\mu$m emission. 
This result implies that mid-IR radiative pumping, or populating, of the
J=1 level of HCN by a mid-IR vibrational transition is not important 
compared with the collisional excitation by dense molecular hydrogen. 
Furthermore, large 
velocity gradient calculations justify the use of HCN J=1-0 emission 
as a tracer of high-density molecular gas 
($\approxgt 3\times 10^4/\tau$~cm$^{-3}$) 
and give an estimate of the mass of dense molecular gas from HCN observations.
Therefore, $L_{\rm HCN}$ may be used as a measure of the total mass
of dense molecular gas, and the luminosity ratio
$L_{\rm HCN}/L_{\rm CO}$ may indicate the fraction 
of molecular gas that is dense.

\end{abstract}


\section{INTRODUCTION}

Though CO traces most of the molecular gas mass in galaxies, it does 
not necessarily trace active star-forming regions where the gas 
density is more than 10 times higher than the average. These regions 
are better traced by high dipole
moment molecules like HCN, CS and HCO$^+$. 
Consequently, these molecules have often been used as
probes of physical conditions in giant molecular cloud (GMC) 
cores --- the regions of active star formation in the Milky Way.
Yet there have been few systematic studies of dense gas in
GMCs in the disk of the Milky Way. Lee, Snell, \& Dickman (1990) 
have found that for Milky
Way disk GMCs, the ratio of CO/CS intensity is 156, much larger than 
for Galactic center clouds and an order of magnitude larger than for 
the archetypal ULIG Arp~220 
(Solomon, Radford, \& Downes 1990).
Some dense clumps in some nearby GMCs have also been surveyed in CS 
(Lada, Bally, \& Stark 1991; Plume \etal 1997) and 
in HCN (Helfer \& Blitz 1997b). Although there are some recent
attempts in Galactic plane HCN, CS, and CO observations 
(Helfer \& Blitz 1997b; McQuinn \etal 2002 and references therein), 
only the central region of the Galaxy 
has been extensively mapped with dense molecular gas 
tracers (\eg, CS, Bally \etal 1987,
1988; HCN, Jackson \etal 1996; Lee 1996). Systematic and
unbiased large-scale observations of the dense molecular gas in 
Galactic disk GMCs need continued study since the
large-scale (kpc) distribution of dense gas in the Milky Way disk
(cf. McQuinn \etal 2002) is a basis for comparison with dense molecular
gas observations in external galaxies (\eg, Paglione \etal 1998). 

HCN and CS are probably the most frequently observed interstellar
molecules after CO.
Because of their higher dipole moments ($\mu _0 \sim$ 2.0 -- 3.0 D),
they require about 2 orders of magnitude higher densities for 
collisional excitation than CO ($\mu_0 \sim$ 0.1 D).
HCN is one of the most abundant high 
dipole moment molecules and traces molecular gas at densities 
$n$(H$_2) \approxgt 3\times 10^4/\tau$~cm$^{-3}$ (compared to 
densities at $\approxgt 500$~cm$^{-3}$ traced by CO). 
HCN J=1-0 emission has been so far detected toward the nuclei
of $\sim$ 10 nearby galaxies (Nguyen-Q-Rieu, Nakai, \& Jackson 1989; 
Nguyen-Q-Rieu \etal 1992; 
Henkel \etal 1990), 10 relatively distant luminous and ultraluminous
infrared galaxies (LIGs and ULIGs with IR luminosities\footnote{The 
total IR luminosity (from 8 $\mu$m to 1000 $\mu$m) is calculated 
according to the prescription in Sanders \& Mirabel (1996) 
using measurements from all 4 \IRAS flux bands.} $L_{\rm IR} > 10^{11}$ \ls~ 
and $>10^{12}$ \ls, respectively) including three less luminous,
``normal'' spiral galaxies (Solomon, Downes, \& Radford 1992), 
12 nearby starburst and/or
normal galaxies (Helfer \& Blitz 1993), 10 interacting galaxies and
infrared galaxy mergers (Aalto \etal 1995), 13 Seyfert galaxies 
(Curran, Aalto, \& Booth 2000), and a few nearby CO-bright galaxies 
(Israel 1992; Sorai \etal 2002; Kuno \etal 2002). There are significant
overlaps in the galaxy samples among these various observations, and
the total number of galaxies
with HCN detections is still $\approxlt 30$, particularly the
total HCN emission measured globally from the galaxies. A recent 
summary of HCN observations in centers of nearby galaxies can
be found in Sorai \etal (2002) and Shibatsuka \etal (2003). 

In a few of the nearest
galaxies, interferometer HCN maps have also been obtained in the 
nuclear regions (Carlstrom \etal 1988; 1990; Radford \etal 1991; 
Downes \etal 1992; Brouillet \& Schilke 1993; 
Jackson \etal 1993; Helfer \& Blitz 1995, 1997a; Tacconi \etal 1994; 
Shen \& Lo 1995; Paglione, Tosaki, \& Jackson 1995b; 
Kohno \etal 1996; Kohno, Kawabe, \& Vila-Vilaro 1999). 
CS emission, another dense molecular gas 
tracer, has been observed in the central regions of $\sim 10$ 
galaxies as well including interferometry imaging 
(Mauersberger \& Henkel 1989; Mauersberger \etal 1989; 
Sage, Shore, \& Solomon \etal 1990; Helfer \& Blitz 1993;
Xie, Young, \& Schloerb 1994; Paglione \etal 1995a; 
Peng \etal 1996; Reynaud \& Downes 1997; Wild \& Eckart 2000).
The presence of large amounts of dense molecular gas 
in the central $\sim$ 1 kpc of galaxies has
been well established (for an earlier review, see Mauersberger \& 
Henkel 1993). 

The main goal of
this paper is to present a systematic HCN survey of a large sample of 
normal spiral galaxies and relatively distant LIGs and ULIGs.
This includes an 
attempt to measure the total HCN emission in nearby large spiral 
galaxies by mapping the HCN distribution along the major axes (Gao 1996,
1997). We examine correlations between HCN emission and various
\IRAS~ bands and discuss the excitation mechanisms of HCN. We here 
argue that HCN is primarily excited by collision with molecular 
hydrogen at high density and thus can be used as a tracer of 
dense molecular gas. We concentrate here on 
presenting observations of HCN J=1-0 only since it is almost always
the strongest line among these high dipole moment molecules, and 
thus the easiest and most often observed dense gas tracer
in external galaxies (Gao 1996). We have not only doubled the total 
number of galaxies with HCN J=1-0 detections in the literature for the last
decade, but we have also extensively mapped the 
inner disk HCN emission in a few nearby galaxies in order to estimate the 
total HCN emission. We will explore in detail the extent 
and radial distribution of dense molecular gas in nearby galaxies 
in a future paper and present the detailed HCN mapping observations
there (Y. Gao \& P.M. Solomon in preparation).
In a companion paper (Gao \& Solomon 2004, Paper II), we fully discuss 
the interpretation and implications of our HCN survey. 

We describe the sample selection and characteristics of our HCN survey
in the next section. In \S~3, we discuss our various 
observations and the observational strategies conducted at different 
telescopes. The survey results are presented in \S~4, where we
show all the HCN J=1-0 and CO spectra, tabulate the
observational data, and summarize 
the global quantities for all galaxies. We have confirmed the tight
correlation between the IR and HCN emission, initially proposed 
by Solomon \etal (1992) based on the HCN observations of a sample
of 10 representative galaxies, 
in a large and statistically significant sample of more than 60 
galaxies when both samples are combined. We have also checked 
for any significant
dependence of the HCN emission upon the excess mid-IR emission
as indicated by an \IRAS flux ratio of the 12 and 100 $\mu$m 
emission or the ratio of the 12 $\mu$m and total IR emission 
to address the possibly enhanced excitation of HCN J=1-0 emission by 
mid-IR radiative pumping through a vibrational transition. Moreover,
the use of HCN as a tracer of
high-density molecular gas is discussed using 
large velocity gradient (LVG) calculations.
Finally, we summarize the main points of this HCN survey in \S~5.

\section{THE HCN SURVEY SAMPLE} 

Our goal is to survey a large sample of galaxies ($> 50$) 
with a wide range of IR luminosities where the total HCN J=1-0 emission
from the whole inner disk or entire galaxy can be measured. 
Essentially, all galaxies with strong 
CO and IR emission were chosen for our HCN observations. 
Our HCN survey sample is mainly drawn from the CO 
observations of Solomon \etal (1997), Sanders, Scoville, \& Soifer (1991), 
Tinney \etal (1990), Solomon \& Sage (1988), and Young \etal (1989, 1995).
Almost all galaxies with an antenna temperature measured by
various telescopes to be greater than 100~mK (or 50~mK for
LIGs) were included as candidate galaxies.
These cutoffs were specifically used in order to select all
truly CO-bright galaxies in the hope of detecting their HCN emission
(which is typically 10--20 times weaker than CO)
with a rms sensitivity at the level of $\sim 1~$mK. 
Most IR-bright galaxies with \ $f_{\rm 60 \mu m} > 50 {\rm Jy, \ or} 
\ f_{\rm 100 \mu m} > 100 {\rm Jy}$ \ 
are included as well in our sample (Rice \etal 1988; 
Strauss \etal 1992). We 
thus selected $\sim 10$ nearest CO and/or IR bright large
galaxies with \ 
$cz \approxlt 1,000~\kms \ {\rm and} \ D_{25} \approxgt 10'$, \ 
and a few tens of relatively distant yet still nearby ``normal'' 
spiral galaxies and LIGs. Moreover, most northern ULIGs in the 
local universe (cz $<$ 20,000 \kms) were also added to 
enlarge our sample at this extremely high IR luminosity. 

We detected HCN J=1-0 emission from 52 galaxies, and only one galaxy
was not detected. Table 1 lists all 53 sources, along with 
some basic properties of the galaxies useful for our observations and 
data analysis. The nearest large galaxies that require mapping to estimate
the total HCN emission are highlighted in boldface in Table~1, 
and the detailed observational results 
of these galaxies will be discussed separately 
(Y. Gao \& P.M. Solomon 2004, in preparation). Seven ULIGs
are marked with a star symbol in Table~1.
Our HCN survey sample also contains essentially a statistically 
complete sample for northern ($\delta \ge -35^{\circ}$) galaxies
with $f_{\rm 100\mu m} \ge 100$ Jy, which includes about 30 
of the brightest \IRAS galaxies.


For the purpose of the detailed statistical analysis to be 
presented in Paper II, we 
included almost all HCN data 
available in the literature where the total HCN emission from
galaxies can be estimated to augment our sample. There are 
about a dozen mostly IR-luminous galaxies with total HCN
emission measured, almost entirely from Solomon \etal (1992),
but M51 and NGC~4945 are from Nguyen-Q-Rieu \etal (1992) and
Henkel \etal (1990), respectively.
Even with these supplements and those intentionally added ULIGs,
the survey sample
is not severely biased toward the high IR luminosity end. 
As shown in Figure~1a, although most of our sample galaxies are
IR bright or the brightest in the local universe, 
more than half have IR luminosities less than $10^{11} \ls$, and among them
most have $L_{\rm IR} < 3\times 10^{10} \ls$. There seems to be a slight 
bias, however, toward high CO luminosity galaxies (Fig.~1b), 
\ie, more CO-luminous or molecular
gas-rich galaxies have been selected.
This bias mainly arises because most normal spiral galaxies selected 
are CO bright and quite rich in molecular gas,
and essentially all LIGs and ULIGs have high CO luminosities.
Nevertheless, the galaxy number distribution 
in HCN luminosity appears to be rather
unbiased with comparable numbers of galaxies with both low and high HCN 
luminosities, and the entire HCN luminosity distribution covers a
large range of almost 3 orders of magnitude (Fig.~1c).


\section{OBSERVATIONS} 

We used the former NRAO\footnote{The National Radio Astronomy
Observatory is operated by Associated Universities, Inc., under
cooperative agreement with the National Science Foundation.} 12m
telescope at Kitt Peak and the IRAM 30m telescope at Pico Veleta
near Granada, Spain, for  most of our  HCN observations. 
The FCRAO\footnote{The Five College Radio Astronomy
Observatory is operated with support from
the National Science Foundation, and with the permission of the
Metropolitan District Commission, Commonwealth of Massachusetts.} 
14m telescope with the 15 element focal-plane array receiver, 
QUARRY (a configuration of three
rows of five beams with a row spacing of $100''$ and a beam
spacing of $50''$), 
was initially used for mapping some of the nearest strong sources. 
We conducted several
observing runs at each telescope from 1993 April to  1995
December. Additional observations were conducted remotely at the NRAO 12m
telescope in 1997 May and June to confirm some weak detections and 
to further improve the signal-to-noise ratios for some 
HCN spectra. The system temperatures 
at the redshifted HCN line frequencies were typically 220, 250, 
and 500 K on average on the antenna temperature scales of $T_R^*$, 
$T_A^*$, and $T_A^*$ for the NRAO, IRAM and FCRAO telescopes, respectively.
We converted the observed antenna temperature to the main-beam 
temperature scale ($T_{\rm mb}$) for all our observations 
at different telescopes. Table 2 lists various telescope beam
efficiencies measured at different telescopes at both the 2.6 and 
3.4 mm wavebands. At $\sim 3$ mm, we adopted the continuum 
sensitivity (the temperature-to-flux conversion for a point source)
on the $T_{\rm mb}$ scale as $S/T_{\rm mb}=4.8$, 29.0, and 
23.5 Jy K$^{-1}$ for the IRAM 30m, NRAO 12m, 
and FCRAO 14m telescopes, respectively. 


At the IRAM 30m, we used two 512 $\times$ 1 MHz filter banks and the 
autocorrelator in its widest bandwidth
configuration as back ends. These back ends were connected to three 
SIS receivers (3mm, 2mm, 1mm) that were usually used to 
simultaneously observe HCN J=1-0, CS (3-2), and CO J=2-1 lines 
respectively. The telescope beam diameters (FWHM) for CO J=1-0 and 
HCN J=1-0 were $22''$ and $27''$, respectively.  The NRAO 12m was 
equipped with  dual-polarization 3mm SIS
receivers. We connected the dual receivers to a pair of $256 \times 2$ MHz
filter banks as well as the two 600 MHz wide hybrid spectrometers. The 
telescope beam diameters (FWHM) for CO J=1-0 and HCN J=1-0 were 
55$''$ and 72$''$, respectively. At the FCRAO 14m, the corresponding 
back ends for the 15 cooled Schottky 
diode mixers of the QUARRY array were 15
spectrometers of $64 \times 5$ MHz bandwidth. 
The telescope beam spacing of $50''$ within each row of QUARRY is about 
the beam size (FWHM) of the 14m telescope at 3 mm. 

The IRAM 30m telescope used a wobbling secondary with a beam throw 
of $\ \pm 4'\ $ to achieve flat baselines. Similarly, a nutating
subreflector with a chop rate of $\sim 1.25$ Hz was used at the NRAO 12m
together with a position switch (the so-called beam-switch plus 
position-switch, or ``BSP mode'').
The beam throws used were in the range of $\ \pm 2'$ to $\pm 3'$ 
(the larger beam throw was used for nearest large galaxies 
with $D_{25} \approxgt 10'$). Thus, the separations between 
the OFF reference positions and
the ON source position are in the range of $\ \pm 4'$ to $\pm 6'$. 
For a couple of the largest galaxies ($D_{25} \approxgt 20'$) 
observed at the 12m, we employed
position switching in azimuth with offsets comparable to 
the expected CO source sizes to ensure that the reference positions
are completely outside of the sources observed. At the FCRAO 14m, 
we used double-position 
switching in azimuth with offsets comparable to the diameters of galaxies.
The QUARRY receivers of the FCRAO 14m telescope were efficient 
in mapping the CO emission in the inner disks of nearby galaxies, 
but the sensitivity achievable limited our extensive mapping 
of the weak HCN emission outside the nuclear regions. No additional
beams except for the central beam and sometimes the adjacent two 
inner beams were detected in HCN with the 14m.
Therefore, most of the HCN mapping was eventually conducted with 
the 12m and 30m telescopes after the initial trial observing runs 
with the QUARRY at the FCRAO 14m.

We mainly used the IRAM 30m to observe most (rather distant) ULIGs 
in the sample and to map some nearby
starburst galaxies of mostly smaller optical diameters, given the 
smaller telescope beam compared with 
the source size and extent for achieving effective mapping. 
Essentially all other observations were conducted
with the NRAO 12m so that one beam measurement covered all emission
from relatively distant galaxies including some interacting/merging 
galaxy pairs. Almost all nearby spiral 
galaxies were mapped along the major axes with the NRAO 12m. 
These 12m observations included both the HCN observations 
initially attempted with the FCRAO 14m QUARRY array, and in several 
cases, the major axis mapping observations from the IRAM 30m in order 
to properly compare the total HCN emission estimated 
with different telescopes (see Table~3). This is essential
to better characterize the radial HCN distribution in galaxies 
(Y. Gao \& P.M. Solomon 2004, in preparation). 

All observed positions were chosen by checking and comparing 
the positions of the peak radio continuum emission
as determined from the VLA snapshot survey (Condon \etal 1990)
or the best positions
(with uncertainties $\approxlt 10''$) then listed in the
NED.\footnote{The NASA/IPAC Extragalactic Database (NED)
is operated by the Jet Propulsion Laboratory, California Institute of
Technology, under contract with the National Aeronautics and Space
Administration.} The observed spectra from all telescopes 
(except for the FCRAO 14m) were smoothed to $\sim$8 MHz resolution, or
$\sim 20 ~\kms$ and $\sim 27 ~\kms$ at 2.6 and 3.4mm, respectively. 
Occasionally, in several of the weakest HCN sources, the spectra
were smoothed to 16 MHz to boost the signal-to-noise ratio. 
Linear baselines were removed from the spectra after emission 
line windows were chosen to estimate the integrated line 
intensities. We converted the line temperatures
to the main-beam brightness temperature scale $T_{\rm mb}$ 
(Tables~2 and 3). 
In many cases, CO lines were also observed from the same 
telescope for comparison and for determination of the HCN
line emission windows. All HCN observations of galaxies so far 
have shown that the significantly detected HCN line profiles 
have the same line widths as the CO lines (\eg, Solomon \etal
1992; Helfer \& Blitz 1993; Gao 1996).  
With the CO spectra obtained from the same positions or available 
in the literature from the same telescope, the integrated line 
intensities of the weakly detected HCN lines can be more accurately
estimated.

Calibrations at all three telescopes were done using the standard 
chopper wheel method (Kutner \& Ulich 1981).
The absolute antenna temperature scale was checked by either
observing the standard calibration sources or by the measured CO line
strengths of the targets
(mostly for the nearest strong sources) as compared with the 
standard spectra at the telescopes and the known CO spectra
available in the literature.
Pointing was checked by observing quasars or planets 
about every 2---3 hours, with typical errors of $\sim 3''\ $ 
for the IRAM 30m, and $\sim 5''$ --- 10$''$ \ for the NRAO 12m and 
the FCRAO 14m. Focus was also monitored a few times per day mostly
using planets. More detailed discussions of the observations and 
the antenna temperature scales 
at different telescopes can be found in Gao (1996, 1997).

\section{RESULTS AND DISCUSSION}

\subsection{Spectra}

We present all HCN spectra in Figs.~2 and 3 for the IRAM 30m 
and the NRAO 12m telescopes, respectively, including a few weak 
($\sim 3 \sigma$) detections. For most galaxies, we also show the CO 
spectra taken at the same position from the same telescope
for comparison. A few CO spectra in Figure~2, however, are 
CO J=2-1 lines, which have much smaller
beam sizes compared with that of the HCN J=1-0 spectra. In some 
cases, the velocity range of CO emission, obtained from 
the published CO spectra in the literature 
with the same telescopes or comparable telescope beam sizes, 
is indicated instead.
NGC~253, NGC~2903, M82, and NGC~6946 were mapped by both 
the IRAM 30m and the NRAO 12m and therefore appear in both Figs.~2 and 3.
Only the central HCN spectra of those nearby galaxies highlighted 
in Table~1 are displayed here; the detailed HCN mapping will be shown 
in a future paper (Y. Gao \& P.M. Solomon 2004, in preparation).



We detected HCN emission in essentially all of the sources 
(except NGC~1055) listed in Table~1, though a few 
galaxies were only marginally detected.
The observed line intensities (of both HCN and CO if applicable)
are listed in Table~3 together with other measured quantities 
useful for analysis. 
The derived global properties of the line luminosities and their ratios
using methods discussed below (\S 4.2) are listed 
in Table~4. 



Fully sampled mapping is essential
in order to estimate the total HCN emission in large
spiral galaxies (Gao 1996, 1997; Y. Gao \& P.M. Solomon 2004, in preparation)
because a substantial fraction of the total HCN emission
is distributed outside the $\sim 1$ kpc nuclear regions (which are
covered by the central beams in these observations). Therefore, 
we did not add more nearby galaxies with HCN
detections from their centers from the literature into our HCN sample. 
Those data only measure lower limits to the HCN luminosity since 
only the central beam detections were made for these
nearby large spiral galaxies, and the HCN emission outside the
central $\sim 1$ kpc nuclear regions is unknown.
Notice that about 15\% of the galaxies in our HCN survey sample 
also have either upper or lower limits in the total HCN emission
or luminosity (Tables~3 \& 4).
The upper limits in the HCN measurements correspond to
weak detections ($\sim 3\sigma$) or non-detections 
($\approxlt 2\sigma$, only for NGC~1055) or to 
nearby galaxies with good detections in the 
center but with only weak detections or non-detections 
at the off-center positions.
The lower limits in the HCN luminosity are for a few nearby 
galaxies with good HCN detections from the galaxy centers, but no 
systematic mapping or detections at off-center positions 
(NGC~2903 is exceptional since it was mapped with the IRAM 30m, 
yet more sensitive and extensive mapping is still required to estimate the
total HCN emission). This is similar to the case of
more than a dozen nearby galaxies with only central beam HCN observations
in the literature (thus lower limits to the total HCN emission). 
We include our own limits, but not the HCN limits 
in the literature in the final HCN sample.

\subsection{HCN and CO Line Luminosities}

For sources much smaller than the telescope beam, the HCN luminosity 
is calculated using 
\begin{equation}
L'_{\rm HCN}=\int T_{\rm b}({\rm HCN})\, dV\, d\Omega_{\rm s}\, d_A\, ^2 
\approx \pi/(4ln2)\, \theta_{\rm mb}\, ^2\, I_{\rm HCN}\, d_L\, ^2 (1+z)^{-3} 
\end{equation}
(see Solomon \etal 1997 for details, $L'$ with
a unit of K~\kms~pc$^2$ is introduced to distinguish it
from normal $L$, which has proper power units), where \
$\Omega_{\rm s}$ is the solid angle subtended by the source area,
\begin{equation}
I_{\rm HCN}=\int T_{\rm mb}({\rm HCN}) dV
=\int T_{\rm b}({\rm HCN})\, dV\, d\Omega_{\rm s}/[(1+z) \Omega_{\rm s*b}]
\end{equation}
is the observed integrated line intensity (as converted to the 
beam-diluted main-beam temperature scale), 
$T_{\rm b}({\rm HCN})$ is the intrinsic source brightness temperature,
$\Omega_{\rm s*b} \approx \pi/(4{\rm ln}2)\, \theta_{\rm mb}$ 
is the solid angle of the source convolved with the diameter (FWHM) 
of the telescope Gaussian beam $\theta_{\rm mb}$,
 and $d_L$ and $d_A$ are the luminosity 
distance\footnote{$H_0$=75\kms~Mpc$^{-1}$ and $q_0=0.5$ were used 
in this paper} and the angular size distance to the source, respectively, 
related by $d_L= d_A (1+z)^2=7994.5[1+z-(1+z)^{1/2}]$~Mpc. 

For large galaxies where HCN was observed at
several positions along the major axis, the total integrated
HCN flux over the source, 
\begin{equation}
F_{\rm HCN}=\int T_{\rm mb}({\rm HCN}) dV d\Omega_{\rm s*b},   
\end{equation}
is estimated following the empirical formula adopted in
Solomon \& Sage (1988; also Sage 1987) to calculate 
the total CO emission. The total HCN emission is obtained by 
summing all the measured line fluxes along 
the major axes with weights determined from considerations of
$\theta_{\rm mb}$, the (source) disk size, $D_{25}$, and
the disk inclination, and assumption of an axisymmetric source
distribution. Thus, for nearby galaxies, 
\begin{equation}
L'_{\rm HCN}=\int T_{\rm mb}({\rm HCN})\, dV\, d\Omega_{\rm s*b}\, 
d_A\, ^2 (1+z) = F_{\rm HCN}\, d_L\, ^2 (1+z)^{-3}. 
\end{equation}

The CO luminosity is calculated similarly with the
same formula listed above (switching HCN with CO). For brevity, 
we hereafter use the symbol $L$ rather than $L'$ for the line 
luminosity in unit of K~\kms~pc$^2$ for both CO and HCN emission.

For those nearby galaxies that we observed with different 
telescopes, the agreement in the estimated total HCN fluxes is 
rather good (Table 3), given the weak HCN lines detected outside the 
nuclear regions in most cases. For all positions observed in HCN, 
the permitted velocity ranges are available either from our own,
or published, CO observations. As discussed in \S~3,
the weak detections or upper limits for 
the HCN line intensities can be better estimated 
when the HCN line emission windows are known/assumed from
the CO lines, and thus the HCN
results are rather robust. The uncertainties in the total
HCN (and occasionally CO) luminosities are up to 
$\sim 50\%$ including the calibration errors of different telescopes,
but in most cases only $\sim 30\%$ or better. The largest 
uncertainties are only present
in a few galaxies with the marginal detections ($\sim 3\sigma$) 
and in some nearby large galaxies with very limited points 
observed where extensive mapping along their major axes is
still required for more accurate estimates of the total HCN emission.

\subsection{Correlation between HCN and IR Emission}

Figure~4 shows a tight linear correlation between $L_{\rm HCN}$ and
$L_{\rm IR}$ over 3 orders of magnitude in IR luminosity. 
The originally proposed correlation, based on 10 representative 
galaxies (Solomon \etal 1992), is also noted for comparison. 
We recalculated \lhcn, \lco~ (see \$4.2) based on the measured line 
intensities $I_{\rm HCN}$ and $I_{\rm CO}$ listed in Table~1 of Solomon
\etal (1992). Obviously, the new HCN data
from a sample more than 5 times larger closely follow the trend
established by Solomon \etal (1992) and have about the same scatters.
A least--squares fit including published data 
(almost entirely from Solomon \etal 1992), yet excluding our HCN limits, 
yields a slope of 1.00 $\pm$ 0.05, with a correlation coefficient
R=0.94 (R$^2 = 0.88$). The correlation implies
$logL_{\rm IR}=1.00logL_{\rm HCN}+2.9$. 
When we include all galaxies in the fit and take the limits given in
Table 4 as the HCN luminosities, the linear 
relation remains almost the same with \ 
$logL_{\rm IR}=0.97logL_{\rm HCN}+3.1$ \ 
and the same correlation coefficient.


We have therefore confirmed the tight correlation between IR and HCN 
luminosities in a large, statistically significant sample. 
LIGs/ULIGs have the highest $L_{\rm HCN}$ and highest ratio
of \lhcn/\lco, and many
of them have $L_{\rm HCN}$ even larger than $L_{\rm CO}$ of
the Milky Way, which implies a relatively high fraction of 
dense molecular gas as well as an extremely large reservoir 
of dense molecular gas (discussed further in \S~4.5).
The detailed discussion of the tight correlation
between $L_{\rm IR}$ and $L_{\rm HCN}$ and other relationships,
as well as their implications, will be fully explored in Paper II. 
In this paper, we would
like to show that HCN emission is indeed a good tracer of
dense molecular gas (in the next two sections). Therefore,
the tight correlation between far-IR and HCN suggests that active 
star formation is strongly related to the mass of dense molecular 
gas and the main source of the strong IR emission in LIGs/ULIGs
is mostly from active star formation. 

In Figure 5, we have divided the sample into two subsamples: 
LIGs and ULIGs with $L_{\rm IR} \ge 10^{11} \ls$ are plotted 
as filled circles,
and less luminous galaxies are shown as open circles. Apparently,
all LIGs/ULIGs have \lhcn~ larger than 10$^8$~\ll, whereas the less
luminous normal spiral galaxies have $\lhcn \approxlt 10^8$~\ll.
We also note that several ULIGs have \lhcn/\lco~ as high as 
$\sim 0.25$, whereas most normal galaxies have \lhcn/\lco~ 
nearly 10 times smaller. Other systematic differences between 
these two subsamples can also be easily seen in various 
other correlations presented in Paper II. 


\subsection{Excitation of HCN}

We here try to empirically demonstrate that the radiative pumping, 
or populating, of the J=1 level of HCN through a vibrational
transition at 14~$\mu$m wavelength, does not enhance
significantly the excitation of the HCN J=1-0 emission.
Since the mid-IR vibrational transition of HCN falls within 
the \IRAS 12~$\mu$m bandpass and there are no other direct 
measurements of the 14~$\mu$m emission available, we use the
12~$\mu$m flux as an approximation. We show in Figure~5 two correlation 
plots that involve the \IRAS 12~$\mu$m emission to illustrate that 
the enhancement in the excitation of HCN through a vibrational
transition, by this potential mid-IR radiative pumping, as postulated 
by Aalto \etal (1995), is not important.
Previous HCN modeling also found that the radiative pumping
from the dust continuum emission
has little effect on the excitation of the rotational transitions
in the vibrational ground state of HCN (e.g., Stutzki \etal 1988;
Paglione, Jackson, \& Ishizuki 1997).

Although there is still a good correlation between
the 12 $\mu$m and HCN emission (a squared correlation 
coefficient of R$^2 = 0.75$), 
the relationship is not particularly 
strong as compared with the much tighter correlation between IR and 
HCN emission (Fig.~4, R$^2 = 0.88$). 
In fact, the correlation between the 12$\mu$m 
and HCN emission appears to be the weakest
among the correlations between the four \IRAS waveband 
and HCN emission. Both the 60 and 100$\mu$m 
luminosities correlate much better with the HCN luminosity. 
For example, the correlation between 100$\mu$m and HCN emission 
(R$^2 = 0.89$) is significantly better than that between 12$\mu$m and 
HCN emission (R$^2 = 0.75$). A three-parameter
fit of HCN emission as a function of both 12$\mu$m and 100$\mu$m
emission clearly shows that most of the contribution to the fit is from
the 100$\mu$m emission alone, whereas there is little from the 12$\mu$m 
emission. Should the mid-IR pumping
be an important effect in the excitation of HCN in the 
vibrational ground state of HCN, a tighter correlation
between the 12$\mu$m and HCN luminosities than any correlations
between the HCN and 60$\mu$m, 100$\mu$m, and total IR luminosities would 
be expected. We observe the contrary.

We can also use the luminosity ratio, rather than the single 
luminosity alone, to normalize the correlations in order to 
eliminate the distance and galaxy size dependencies inherent 
in the luminosity correlations (more examples are in Paper II). 
There is only a weak correlation 
between $L_{\rm 12\mu m}/\lco$ and \lhcn/\lco~(R$^2 = 0.29$), 
whereas the correlation between $L_{\rm 100\mu m}/\lco$
and \lhcn/\lco~ is still fairly strong (R$^2 = 0.56$).
Clearly, these results suggest that the mid-IR emission
is only loosely correlated with the HCN emission, and the correlation
between the far-IR emission and the HCN emission is significantly
tighter in comparison.
This essentially rules out the possibility that the mid-IR 
pumping in the vibrational 
ground state of HCN could be one of the significant contributors to
the HCN J=1-0 emissivity.

To strengthen these results, we compare the HCN luminosity
and the ratio of HCN and CO luminosities (\lhcn/\lco) with the 
excess 12~$\mu$m emission estimated by either the ratio of
12$\mu$m and total IR or the ratio of 12$\mu$m and 100$\mu$m
(or 12$\mu$m and 60$\mu$m) luminosities. We plot the luminosity 
ratio of $L_{\rm 12\mu m}/\lir$ versus \lhcn~ in Figure~5a and
the luminosity ratios of \lhcn/\lco~ versus $L_{\rm 12\mu m}/L_{\rm 100\mu m}$ 
in Figure~5b. Obviously, no correlations
are observed in these plots and there is no sign of increased HCN emission
or an increased \lhcn/\lco~ ratio for galaxies with a 12~$\mu$m emission
excess. Similar plots for relationships
between the luminosity ratios of \lhcn/\lco~ and $L_{\rm 12\mu m}/\lir$ 
or between \lhcn/\lco~ and $L_{\rm 12\mu m}/L_{\rm 60\mu m}$ 
show no correlations either. Therefore,
there is little evidence for a direct connection between the 
12~$\mu$m and HCN emission, even for galaxies with excessive 12~$\mu$m
emission (Fig.~5). The contribution of mid-IR radiative pumping to 
the HCN excitation 
must be negligible when it is compared to the collisional excitation
by dense molecular hydrogen (see \S~4.5).

Gao \& Solomon (Paper II) show that 
\lhcn~ can be quite accurately predicted from the IR (or far-IR)
and CO luminosities. In this three-parameter model fit, \lir~
appears to be slightly more important, but \lco~ is also a significant
predictor for \lhcn [the HCN(\lir, \lco) model (see \S~3.3.1 and
Table~3 in Paper II)]. Here we can add 
$L_{\rm 12\mu m}$ as an additional parameter into the 
model fit for \lhcn~ to check whether $L_{\rm 12\mu m}$
is a non-negligible predictor for \lhcn~ as well. This 
could test the significance 
of the mid-IR emission contribution to HCN as compared with that 
of the far-IR and CO emission. We find that the model fit changes 
only marginally with no improvement in the 
correlation coefficient or the rms scatter of the fit. In fact, 
no difference is noticeable in the model fit whether 
$L_{\rm 12\mu m}$ is added or not. Therefore,
even though there exists a correlation between 12$\mu$m and HCN
emission, the correlation is significantly worse than the correlations
between the HCN and IR emission of three other \IRAS bands, 
and the contribution to HCN from 12~$\mu$m emission is
negligible. We conclude that there is no clear 
evidence to claim that the mid-IR pumping can significantly
enhance the HCN emission.

For the archetypal AGN/starburst hybrid galaxy NGC~1068, which has
the strongest mid-IR emission in our HCN sample, the \lhcn/\lco~ ratio
is also one of the highest. In particular, a very high HCN/CO ratio 
($\approxgt 0.3$) within $\sim 100$~pc nuclear region is observed
(Jackson \etal 1993; Tacconi \etal 1994; Helfer \& Blitz 1995) in
the interferometric maps, which appears to be the mid-IR compact
core around AGN (Le Floc'h \etal 2001). Is this an example that 
the mid-IR pumping 
has significantly enhanced the global HCN emission? Apparently not,
since the innermost $\approxlt $100~pc nuclear region around the AGN 
contributes little to the average HCN/CO ratios
in the circumnuclear ($\sim $0.5--1~kpc) starburst ring where
most of molecular gas is located. Also in NGC~1068, 
the ratio $f_{\rm 12\mu m}/f_{\rm 100\mu m} = 0.16$ is much higher 
than in any other galaxy in the sample. Yet, it is
still not certain whether the extreme
excess of the mid-IR emission contributes significantly to
the excitation of HCN. This is because galaxies with the lowest 
$L_{\rm 12\mu m}/\lir$ ratios can have even larger HCN luminosities
than NGC~1068 (Fig.~5a) and galaxies with the lowest
$f_{\rm 12\mu m}/f_{\rm 100\mu m}$ ratios appear to have similar
or even higher \lhcn/\lco~ ratios than NGC~1068 (Fig.~5b). 
Thus, globally the high \lhcn/\lco~ ratio and high \lhcn~ in galaxies
may have nothing to do with the mid-IR emission excess. Again, there 
is no indication that the strong mid-IR emission is linked to 
the high ratio of HCN/CO or large HCN luminosity even for 
galaxies with the most extreme mid-IR emission excesses.

\subsection{Mass of Dense Molecular Gas Traced by HCN}

The mass of molecular gas in galaxies is typically derived from
the CO luminosity using the ``standard'' CO-to-H$_2$ conversion factor 
\begin{equation}
M(H_2)=4.78 \ L_{\rm CO} \ \ms (\ll)^{-1}
\end{equation}
(\eg, Solomon \& Barrett 1991; Young \& Scoville 1991 and references
therein), estimated from the Galactic disk
GMCs in the Milky Way. Although this conversion corresponds to 
$N(H_2)/I_{\rm CO}=3\times 10^{20}$ cm$^{-2}$ (K \kms)$^{-1}$,
rather than the recent calibration of 
$N(H_2)/I_{\rm CO}=1.7\times 10^{20}$ cm$^{-2}$ (K \kms)$^{-1}$
(Dame \etal 2001; Hunter \etal 1997), we use it simply for easy
comparison with previous CO studies of galaxies.
As discussed in Downes, Solomon, \& Radford (1993), 
Solomon \etal (1997), and Downes \& Solomon (1998),
however, we caution that
it might overestimate the molecular gas mass in ULIGs by 
factors of 3--5. Unlike normal spiral galaxies, where a large fraction 
of CO emission comes mainly from GMCs distributed in the inner disk,
the CO distribution in ULIGs is extremely 
concentrated in the nuclear regions and may fill the entire volume
since even the intercloud medium may be molecular (Solomon \etal 1997;
Downes \& Solomon 1998). High-resolution CO imaging of ULIGs has 
indeed shown extremely high concentration of CO emission
in the central few hundred parsecs (\eg, Downes \& Solomon 1998).
For ULIGs, 
the conversion factor is thus lower than 
for normal spiral galaxies. Nevertheless, we still adopt the standard 
CO-to-H$_2$ gas mass conversion (eq. [5]) for convenience
since 90\% of the galaxies
in our sample are not ULIGs. The ratio of
\lir/M(H$_2$), often referred to as the star formation efficiency,
is also listed in Table~4.

Observations of H$^{13}$CN (Nguyen-Q-Rieu \etal 1992; Wild \etal 
1992) indicate that extragalactic
HCN emission is also optically thick, just like CO emission. 
The conversion factor between HCN and the mass of the dense molecular
gas, however, is poorly constrained as there is no direct 
calibration from GMCs or GMC core regions in the disk of the Milky Way. 
From observations of higher transition CS emission, Plume \etal (1997) 
find a typical GMC core has a diameter of 1~pc and mass of 3800\ms.
Then the rms velocity of a virialized cloud with these parameters is 
$\sim 5$~\kms, consistent with the observed FWHM line width.
We have carried out large velocity gradient (LVG) calculations that
include radiative trapping for HCN J=1-0 to determine the mass of 
dense molecular gas from the measured HCN brightness temperatures 
(\eg, Kwan \& Scoville 1975), using the collision rates given by 
Green \& Thaddeus (1974). For an HCN abundance (relative to H$_2$)
of $2\times 10^{-8}$ (\eg, Irvine, Goldsmith, \& Hjalmarson 1987; 
Lahuis \& van Dishoeck 2000) and 
a velocity gradient of $\sim 5$\kms~pc$^{-1}$, this gives
[HCN]/$(dv/dr) \sim 4 \times 10^{-9} (\kms$~pc$^{-1})^{-1}$.
For a kinetic temperature of 20--50~K and the intrinsic HCN line 
brightness temperature $T_{\rm b} > $ 5 K, this gives 
$N(H_2)/I_{\rm HCN} \la 
2\times 10^{21}$ cm$^{-2}$ (K \kms)$^{-1}$
or
\begin{equation}
 M_{\rm dense}({\rm H}_2) \la 25 L_{\rm HCN}\ \ms (\ll)^{-1}
\end{equation}
for H$_2$ at a density ~$\approxgt 10^4/\tau$~cm$^{-3}$, 
where $M_{\rm dense}$(H$_2$) means the dense molecular gas 
as traced by HCN emission, and $\tau \approxgt 1$ is the optical 
depth of the HCN J=1-0 line. 

For higher intrinsic HCN line brightness temperature $T_{\rm b} > 15$~K,
higher gas density n(H$_2) \approxgt 3\times 10^4/\tau$~cm$^{-3}$, 
and the same abundance and
velocity gradient, however, LVG results give $N(H_2)/I_{\rm HCN} \la 
1.3\times 10^{21}$ cm$^{-2}$ (K \kms)$^{-1}$, or
\begin{equation}
 M_{\rm dense}({\rm H}_2) \la 15 L_{\rm HCN}~ \ms (\ll)^{-1}.
\end{equation}

Alternatively, for a virialized cloud core of an average density 
\nhtwo$=3\times 10^4$ cm$^{-3}$ 
and $T_b \sim $ 35 K (\eg, Radford, Solomon, \& Downes 1991),
\begin{equation}
 M_{\rm dense}({\rm H}_2)\approx 2.1 {{\nhtwo^{1/2}}\over{T_b}} L_{\rm HCN} 
\sim 10~L_{\rm HCN}~ \ms (\ll)^{-1}, 
\end{equation}
almost same as that obtained from LVG approximation (eq. [7]).

The LVG calculations above give perhaps an upper limit for
the conversion factor between HCN and the mass of dense molecular
gas (cf. Mauersberger \& Henkel 1993) since the much 
lower intrinsic HCN line brightness temperature $T_b>5$ or 
15~K was used besides other approximations in the LVG estimates.
Higher $T_{\rm b}$ indeed tends to result in a smaller conversion
factor allowed by LVG (eq. [7]). HCN hot cores in regions of massive
star formation usually have extremely high temperature up to hundreds 
of degrees (e.g., Boonman \etal 2001). Thus, it is likely that a much
smaller conversion factor holds under these extreme conditions. 
Moreover, there are some other mechanisms possible to help excite 
HCN molecule, \eg, excess HCN abundance (Bergin, Snell, \& Goldsmith 1996; 
Lahuis \& van Dishoeck 2000) resulting from the active star-forming 
environment and high supernovae events, and collisions with electrons. 
Although these are most likely insignificant (\eg, mid-IR radiative
pumping discussed in previous section) and the dominant mechanism should still 
be collisions with molecular hydrogen, they could work collectively 
to lower the conversion factor given in equations (6) and (7). 
Other constraints such as the total 
molecular gas mass estimated from the CO observations, the higher 
frequency CO line observations, and the dynamical 
mass estimations (\eg, Solomon \etal 1997; Downes \& Solomon 1998), 
all seem to point to a smaller conversion factor, particularly 
for LIGs/ULIGs. Thus, a lower conversion factor given
in equation (8) is more likely.

In fact, since the luminosity ratio \lhcn/\lco~ in ULIGs can be as
large as $\sim 25 \%$ (\eg, Mrk~231, Mrk~273 and \IRAS17208-0014),
the true conversion factor between HCN and the mass of dense 
molecular gas should obviously be much smaller than the upper 
limit allowed by the LVG calculation (eq. [7]) in order to 
reconcile with the total molecular gas mass estimated from 
the standard CO-to-H$_2$ conversion (eq. [5]). Particularly
for ULIGs, even equation (8) might still overestimate the dense
molecular gas mass since $T_{\rm b}$ could be indeed very high 
($\approxgt 50$~K) and the standard
CO-to-H$_2$ conversion of equation (5) should be reduced by 
a factor of 5 in ULIGs (Downes \& Solomon 1998). The main difference
between normal spiral galaxies and ULIGs in the HCN-to-H$_2$ 
conversion, however, is the drastically different molecular line 
brightness temperatures. While $T_b \sim 10$~K is valid for most normal
spiral galaxies, a factor of 5 higher ~$T_b \sim 50$~K~ was
deduced for ULIGs undergoing extreme starbursts (Downes \& 
Solomon 1998), which argues for a much lower HCN-to-H$_2$ 
conversion factor than those given in the above equations.

Nevertheless, this also indicates that ULIGs with the highest 
\lhcn/\lco~ ratio are likely to have most ($\approxgt 50 \%$) 
of their molecular gas at high density ($\approxgt 3\times 10^4/\tau$ cm$^{-3}$) 
and that \lhcn/\lco~ ratio can indeed be used as
an indicator of the fraction of dense molecular gas in galaxies.
On the other hand, \lhcn/\lco~ ratio also indicates the
average molecular gas density in galaxies. Given the large
uncertainty of the HCN-to-H$_2$ conversion, the same as is true for
CO-to-H$_2$ conversion, we suggest that a conversion factor 
of $M_{\rm dense}(H_2)/L_{\rm HCN} \sim 10 \ \ms (\ll)^{-1}$
(eq. [8]) is the more reasonable one to use, and 
we adopt this relation to estimate the total dense molecular 
gas mass in galaxies
as tabulated for \lir/$M_{\rm dense}(H_2)$ in Table~4. Accurate 
determination of the conversion factor between HCN and 
the mass of dense molecular gas needs further extensive studies 
including higher frequency HCN observations (\eg, Jackson \etal 1995;
Paglione \etal 1997) and more detailed modeling.

\section{CONCLUSIONS}

We present systematic HCN J=1-0 observations (complemented
with CO observations) of a sample of  53 IR/CO-bright and/or 
luminous galaxies, which is the largest HCN sample and most 
sensitive HCN survey of external galaxies so far. 
Ultraluminous infrared galaxies have the highest HCN luminosities, 
usually several times larger than the CO luminosity of our own Galaxy. 
Many luminous infrared galaxies have HCN 
luminosities comparable to the CO luminosity of the Galaxy. 
The ratio of HCN and CO luminosities can 
be as large as $\sim 25 \%$ in ultraluminous infrared galaxies, 
nearly an order of magnitude higher than most normal spiral galaxies.

We compared the HCN line emission with the far-IR emission 
(an indicator of the rate of high-mass OB star formation).
All galaxies surveyed 
follow the tight correlation between IR and HCN luminosities
initially proposed by Solomon \etal (1992). Since stars form
only in dense clouds and the HCN-emitting gas (i.e., the
dense molecular gas) and IR emission appear closely and
physically related, then active star formation is the dominant
source of IR emission in luminous and ultraluminous galaxies.
This is fully discussed in Paper II. 

There is no particularly strong correlation between 
the 12 $\mu$m and HCN luminosities. The correlations between
100 $\mu$m (and 60 $\mu$m) and HCN emission 
are much better correlated in comparison. Galaxies with 
excess 12 $\mu$m emission
do not show stronger HCN emission or higher HCN/CO luminosity 
ratios. Thus, mid-IR radiative pumping of HCN excitation
is of little significance when compared with the collisional
excitation by dense molecular gas. 

We also discussed the use of HCN J=1-0 emission as a tracer of 
higher density molecular gas ($\approxgt 3\times 10^4/\tau$ cm$^{-3}$) and
adopted a HCN-to-H$_2$ conversion factor 
$M_{\rm dense}(H_2)/L_{\rm HCN} \sim 10 \ \ms$~(K~\kms~pc$^2)^{-1}$.
Luminous and ultraluminous infrared galaxies usually have much 
higher HCN brightness temperature, which
might result in a lower conversion factor.
Analyses including LVG calculations indicate that the HCN luminosity 
measures the mass of dense molecular gas and the 
HCN/CO luminosity ratio indicates 
the fraction of molecular gas in a dense phase.

\acknowledgments

We thank Simon Radford, Dennis Downes, and Mark Heyer for 
help with some of the observations.
We appreciate the generous support and allocation of observing time
from the NRAO 12m, the IRAM 30m, and the FCRAO 14m.
We would also like to acknowledge Edwin Bergin, Paul Goldsmith,
and Ron Snell for providing their LVG code. The careful review
and helpful comments of the anonymous referee are greatly
acknowledged. This research has made use of the NASA/IPAC 
Extragalactic Database (NED) which is operated by the
Jet Propulsion Laboratory, Caltech, under contract with the National
Aeronautics and Space Administration.

\newpage
\begin{figure}
\plotone{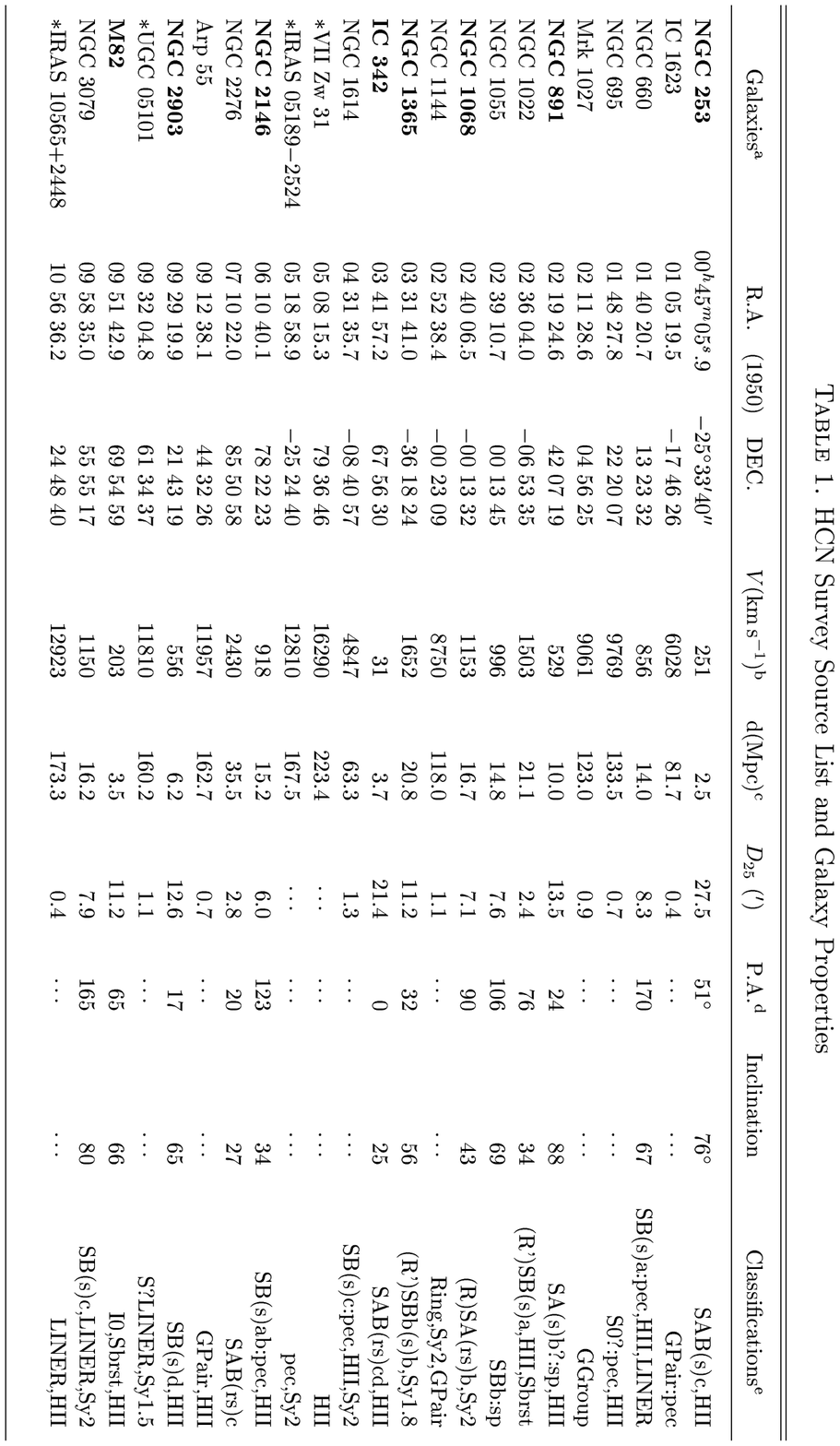}
\end{figure}
\begin{figure}
\plotone{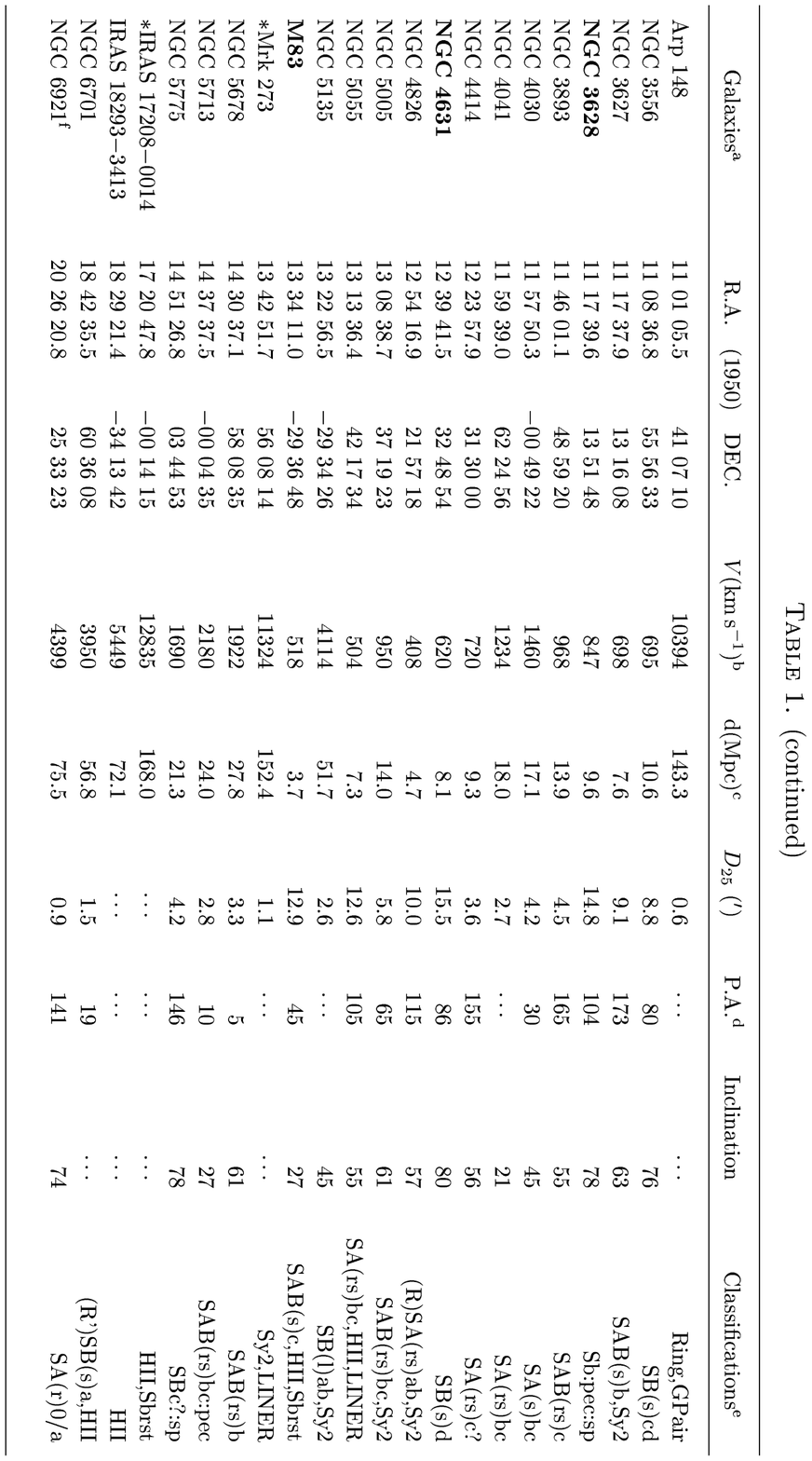}
\end{figure}
\begin{figure}
\plotone{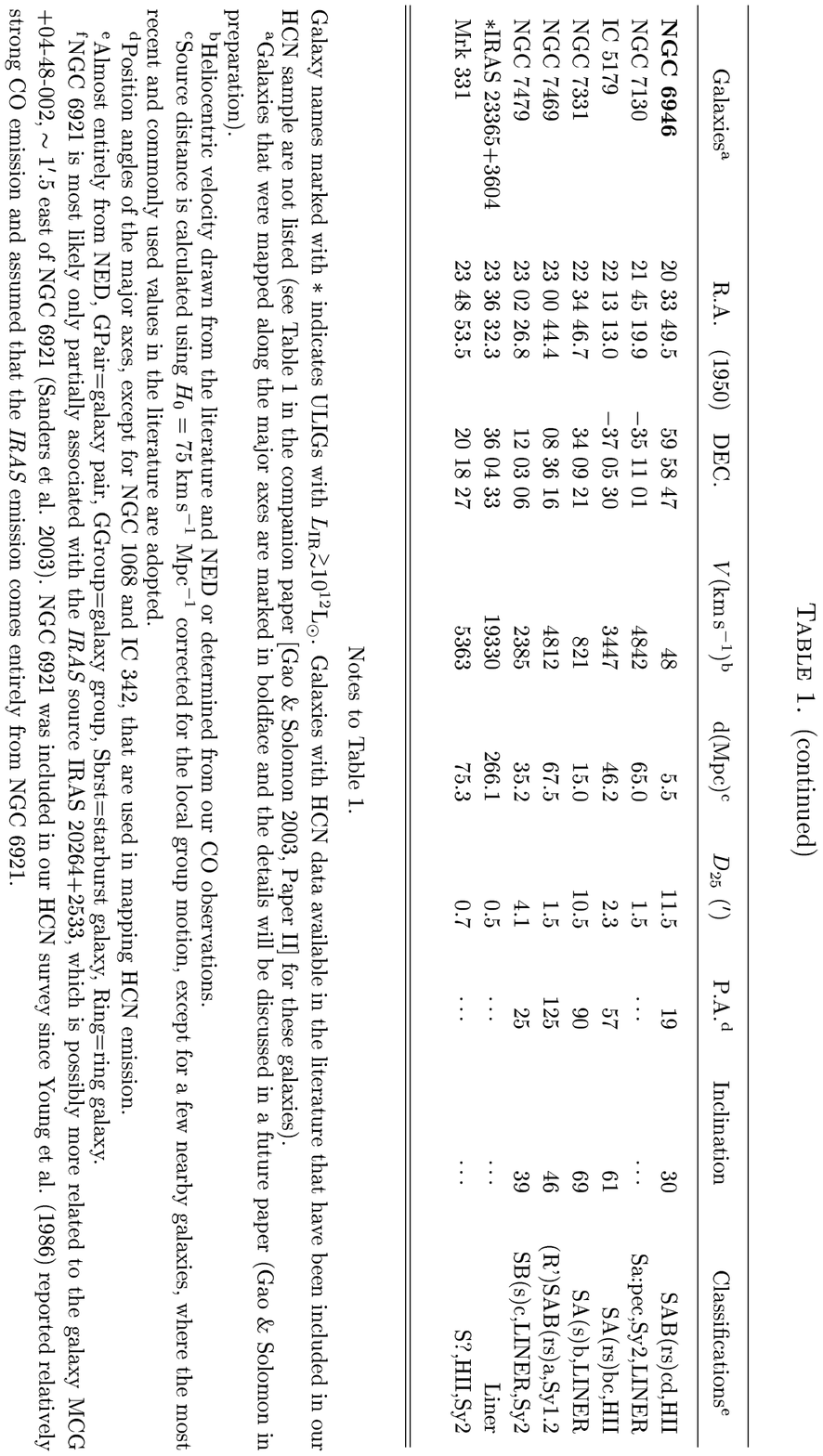}
\end{figure}

\begin{deluxetable}{lccrrrr}
\tablenum{2}
\tablecaption{Telescope Efficiencies at 3 mm\tablenotemark{a}\label{tbl-2}} 
\tablehead{
\colhead{Telescope}      &     \colhead{Typical $T_{\rm sys}$}  &
\colhead{$T$ Scale}  & \colhead{$\eta_{\rm M}^*$\tablenotemark{b}} &
\colhead{$F_{\rm eff}$\tablenotemark{c}}    & 
\colhead{$B_{\rm eff}$\tablenotemark{c}}    &
\colhead{$\eta_{\rm B}$\tablenotemark{d}}
}
\startdata

NRAO 12m & 220 (250)K  &  $T_{\rm R}^*$ & 0.89 (0.83) & & & \nl
IRAM 30m & 250 (270)K  &  $T_{\rm A}^*$ &  & 0.92 (0.92) & 0.75 (0.65) & \nl
FCRAO 14m & 500 (850)K &  $T_{\rm A}^*$ &  &  &  & 0.60 (0.55) \nl
\tablenotetext{a}{Both values at 3.4 mm for HCN J=1-0 and at 2.6 mm for
CO J=1-0 (in parentheses) are given.
These efficiencies were adopted for conversion from the measured
antenna temperature scale to the main-beam $T_{\rm mb}$ scale.}
\tablenotetext{b}{$T_{\rm mb}=T_{\rm R}^*/\eta_{\rm M}^*$, where
$\eta_{\rm M}^*$ is the corrected main-beam efficiency.}
\tablenotetext{c}{$T_{\rm mb}=T_{\rm A}^*(F_{\rm eff}/B_{\rm eff})$, where 
$F_{\rm eff}$ and $B_{\rm eff}$ are the forward and effective beam 
efficiencies respectively.} 
\tablenotetext{d}{$T_{\rm mb}=T_{\rm A}^*/\eta_{\rm B}$, where 
$\eta_{\rm B}=B_{\rm eff}/F_{\rm eff}$ is the main-beam efficiency.}
\enddata

\end{deluxetable}

\newpage

\begin{figure}
\plotone{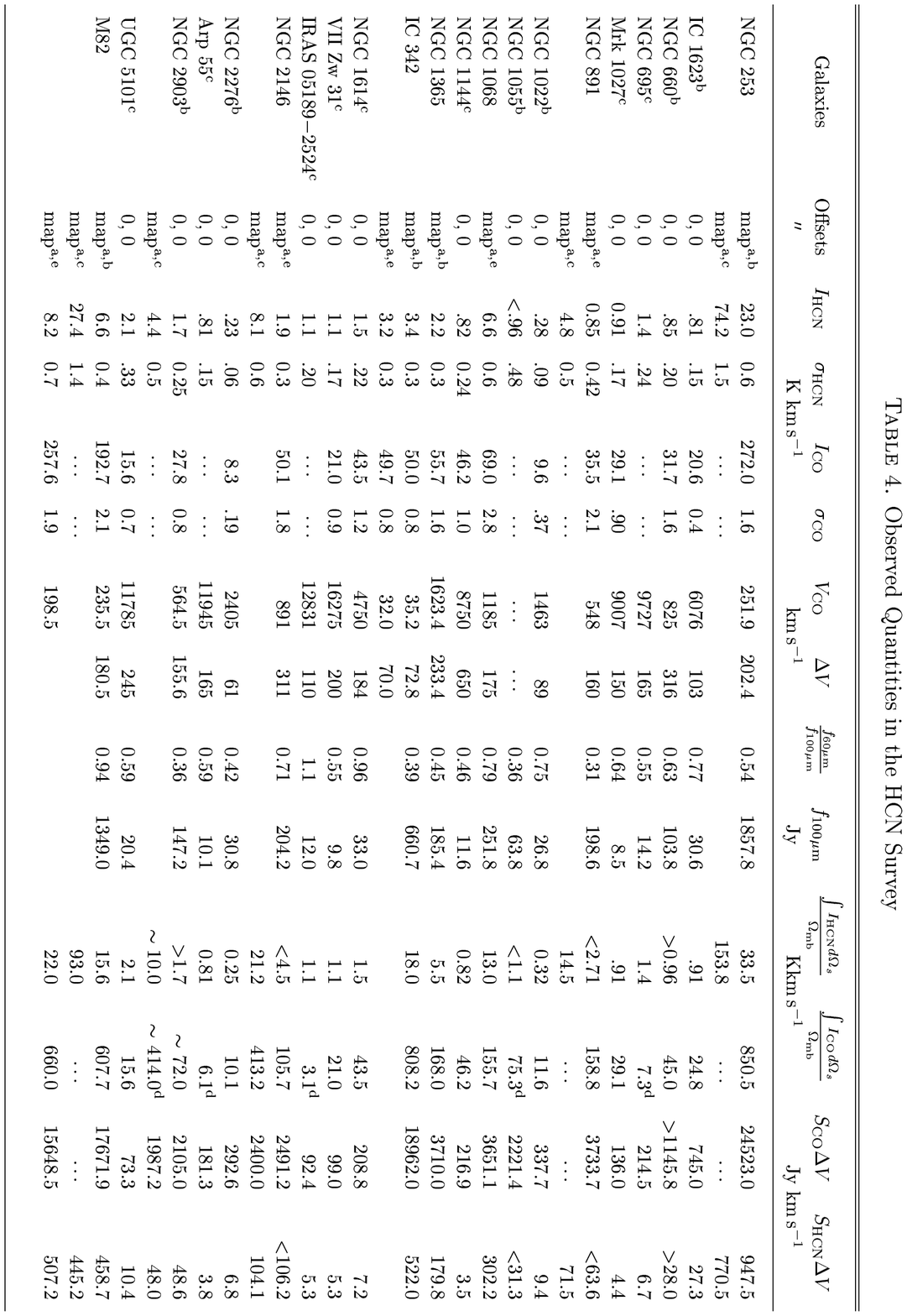}
\end{figure}
\begin{figure}
\plotone{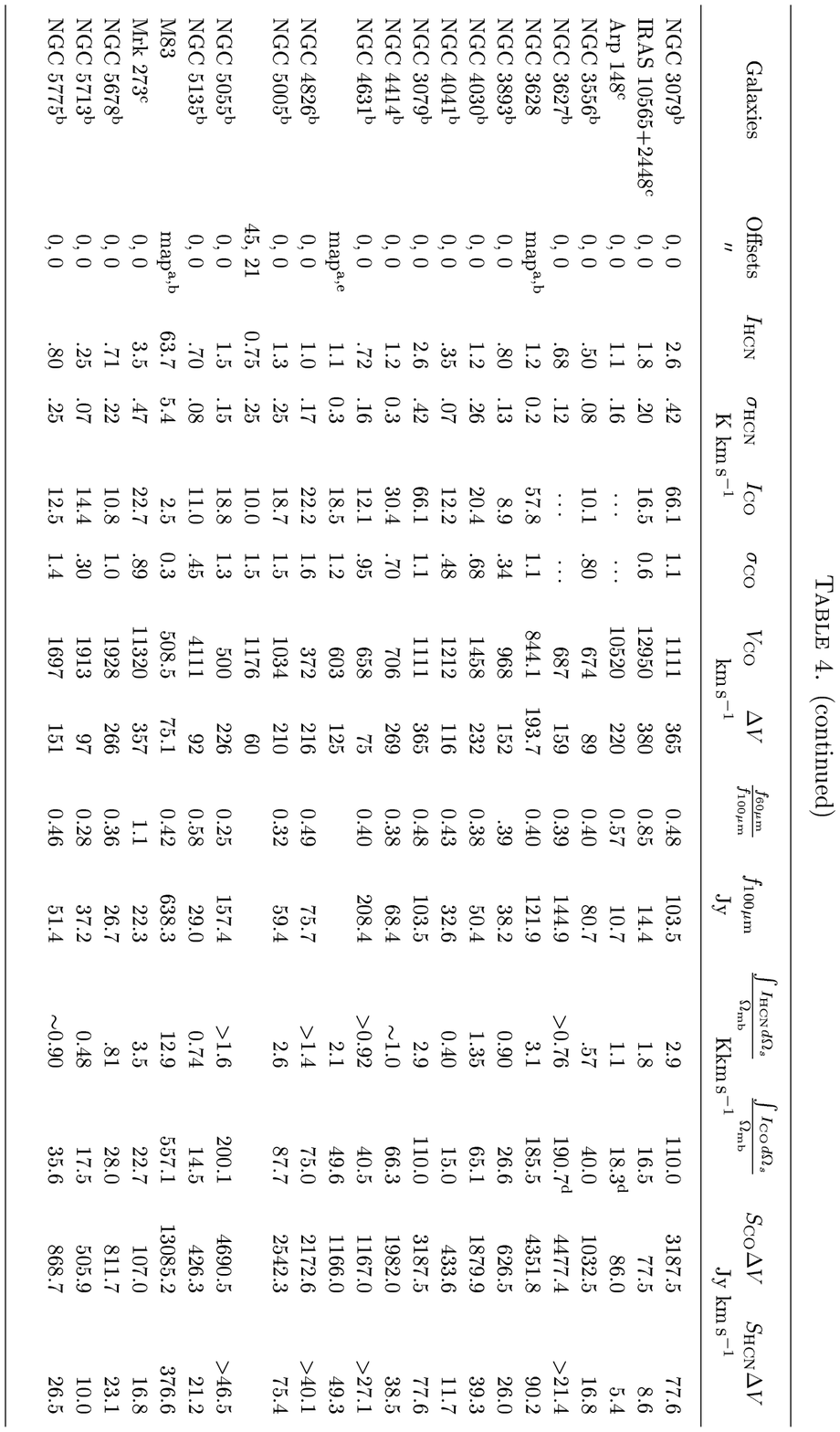}
\end{figure}
\begin{figure}
\plotone{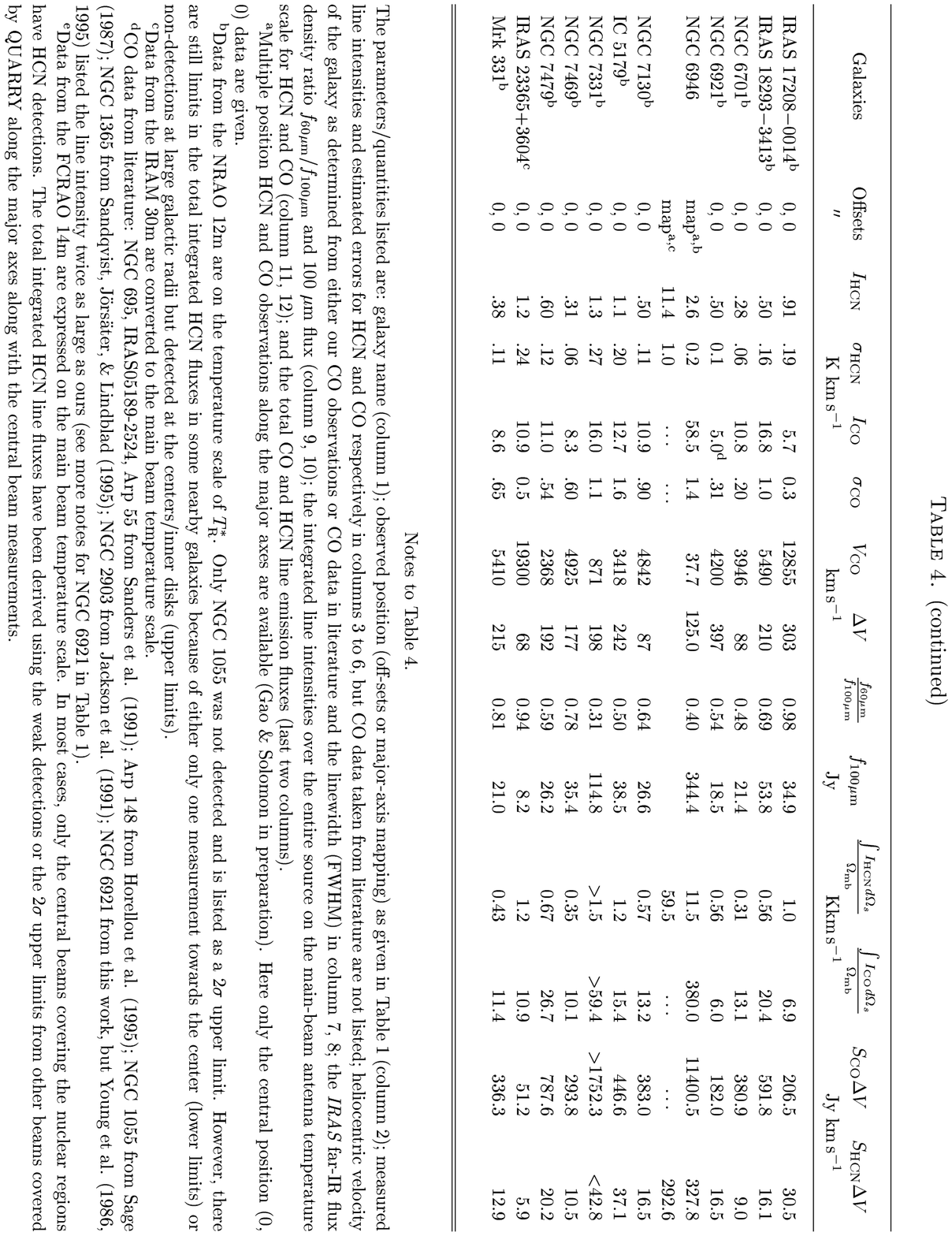}
\end{figure}

\begin{deluxetable}{lrrrrrr}
\tablenum{4}
\tablecolumns{7}
\tablecaption{Global Properties of Galaxies in the HCN Survey \label{tbl-4}}
\tablehead{
\colhead{Galaxies}             &     
\colhead{$L_{\rm IR}$}         &       \colhead{$L_{\rm CO}$}    &
\colhead{$L_{\rm HCN}$\tablenotemark{a}}    &   
\colhead{$L_{\rm HCN}/L_{\rm CO}$}          &
\colhead{${L_{\rm IR}}\over{M_{\rm dense}(H_2)}$\tablenotemark{b}} & 
\colhead{${L_{\rm IR}}\over{M({\rm H_2})}$} \\
\colhead{   }                  & 
\colhead{$10^{10}$ \ls}        &  
\multicolumn{2}{c}{$10^8~\ll$} &   
\colhead{   }                  &
\multicolumn{2}{c}{\ls/\ms}
}

\startdata

     NGC 253   &  2.1 &    4.6 &    0.27 &    0.059 &    77.8 &     9.6 \nl
 IC 1623  & 46.7 &  130.5 &     8.5 &    0.065 &    55.0 &     7.5 \nl
     NGC 660   &  3.7 &    7.3 & $>$0.26 & $>$0.036 & $<$142.3&    10.6 \nl 
 NGC 695  & 46.6 &   92.9 &     4.3 &    0.046 &   108.4 &    10.5 \nl
 MRK 1027 & 25.7 &   41.7 &    1.89 &    0.045 &   136.0 &    12.9 \nl
     NGC 891   &  2.6 &   11.0 &    0.25 &    0.024 &   104.0 &     5.4 \nl
     NGC 1022  &  2.6 &    4.2 &    0.20 &    0.047 &   131.5 &    13.1 \nl
     NGC 1055  &  2.1 &   13.3 & $<$0.37 & $<$0.028 & $>$56.9 &     3.3 \nl
 NGC 1068 & 28.3 &   20.7 &    3.61 &    0.174 &    78.5 &    28.6 \nl
 NGC 1144 & 25.1 &  108.9 &    2.67 &    0.025 &    94.0 &     4.8 \nl
 NGC 1365 & 12.9 &   58.7 &    3.10 &    0.053 &    41.6 &     4.6 \nl
     IC 342    &  1.4 &    9.4 &    0.47 &    0.050 &    30.2 &     3.4 \nl
 NGC 1614 & 38.6 &   24.5 &    1.25 &    0.051 &   306.8 &    33.2 \nl
 *VIIZw31 & 87.1 &  125.0 &     9.8 &    0.078 &    88.9 &    14.6 \nl
 *05189$-$2524 &118.1& 67.0&     6.2 &    0.093 &   190.4 &    36.9 \nl
     NGC 2146  & 10.0 &   12.5 &    0.96 &    0.071 &   122.4 &    16.8 \nl
     NGC 2276  &  6.2 &   10.2 &    0.40 &    0.039 &   156.0 &    12.8 \nl 
 ARP 55   & 45.7 &  125.0 &     3.8 &    0.030 &   120.2 &     7.6 \nl
     NGC 2903  & 0.83 &    2.3 & $>$0.09 & $>$0.036 &$<$104.0 &     7.6 \nl
 *UGC 05101 & 89.2&   50.8 &    10.0 &    0.197 &    89.2 &    36.7 \nl
     M82       &  4.6 &    5.7 &    0.30 &    0.053 &   153.3 &    16.9 \nl
     NGC 3079  &  4.3 &   24.0 &$\sim$1.0&$\sim$0.042&$\sim$43.5&   3.7 \nl
 *10566+2448 & 93.8&   61.5&    10.2 &    0.166 &    92.0 &    32.0 \nl
 ARP 148  & 36.5 & $>$47.0&     4.0 & $<$0.085 &    91.2 & $<$16.2 \nl
     NGC 3556  & 1.35 & $>$4.5 & $>$0.09 &$\sim$0.020& $<$155.0& $<$6.3 \nl 
     NGC 3627  & 1.26 &    4.4 & $>$0.08 & $>$0.017 & $<$160.0&     6.0 \nl
     NGC 3628  & 1.01 &    7.1 &    0.24 &    0.034 &    43.5 &     3.3 \nl
     NGC 3893  & 1.15 &    4.1 &    0.23 &    0.056 &    50.4 &     5.9 \nl
     NGC 4030  & 2.14 &   15.2 &    0.54 &    0.036 &    40.0 &     3.0 \nl
     NGC 4041  & 1.70 &    3.9 &    0.18 &    0.046 &    94.8 &     9.1 \nl 
     NGC 4414  & 0.81 &    4.6 &    0.16 &    0.033 &    54.8 &     3.7 \nl
     NGC 4631  &  2.0 &    2.3&$\sim$0.09&$\sim$0.039&$\sim$222.2& 18.5 \nl
     NGC 4826  & 0.26 &    1.3 & $>$0.04 & $>$0.030 & $<$65.5 &     4.1 \nl
     NGC 5005  &  1.4 &    8.2 &    0.41 &    0.049 &    21.5 &     2.2 \nl
     NGC 5055  &  1.1 &    8.6 & $>$0.10 & $>$0.012 &$<$111.0 &     2.7 \nl
 NGC 5135 & 13.8 &   31.3 &    2.73 &    0.087 &    50.6 &     9.1 \nl 
     M83       &  1.4 &    8.1 &    0.35 &    0.043 &    40.5 &     3.7 \nl
 *MRK 273 &129.9 &   65.0 &    15.2 &    0.234 &    85.5 &    41.8 \nl
     NGC 5678  &  3.0 &   17.2 &    0.75 &    0.044 &    40.0 &     3.6 \nl
     NGC 5713  &  4.2 &    8.1 &    0.22 &    0.027 &   188.9 &    10.7 \nl
     NGC 5775  &  3.8 &   10.9 &    0.57 &    0.052 &    66.7 &     7.2 \nl
 *17208$-$0014 &234.5& 146.9&   37.6 &    0.256 &    62.4 &    33.4 \nl 
 18293$-$3413 & 53.7 &  85.5&   4.03 &    0.047 &    66.6 &    13.2 \nl 
 NGC 6701 & 11.2 &   34.0 &    1.38 &    0.041 &    40.6 &     6.9 \nl
 NGC 6921 & 11.4 &   17.5 &$\sim$2.81&$\sim$0.160&$\sim$41.0&  13.6 \nl
     NGC 6946  &  1.6 &    9.2 &    0.49 &    0.053 &    33.5 &     3.6 \nl
 NGC 7130 & 21.4 &   44.9 &    3.27 &    0.071 &    65.4 &    10.0 \nl
 IC 5179  & 14.1 &$\sim$26.4&  3.42 &$\sim$0.129&   41.2&$\sim$11.2\nl
     NGC 7331  &  3.5 & $>$10.7& $>$0.44 &$\sim$0.041& $<$79.8&  $<$6.8 \nl
 NGC 7469 & 40.7 &   37.1 &    2.19 &    0.059 &   185.4 &    23.0 \nl
     NGC 7479  &  7.4 &   26.7 &    1.12 &    0.042 &    66.1 &     5.8 \nl
 *23365+3604 & 142.0&  85.0 &    15.0 &    0.176 &    94.6 &    35.0 \nl
 MRK 331  & 26.9 &   52.1 &    3.35 &    0.064 &    80.3 &    10.7 \nl 

\tablenotetext{a}{The 2~$\sigma$ 
upper limit is listed for NGC~1055. The lower limits are 
for nearby galaxies where we only detected HCN in the galaxy 
centers or extensive mapping is still required in order to estimate 
the total HCN and CO luminosities.}
\tablenotetext{b}{$M_{\rm dense}=10\lhcn$~\ms/\ll and 
$M(H_2)=4.78\lco$~\ms/\ll, see \S4.5.}
\tablecomments{ULIGs with $L_{\rm IR} \approxgt 10^{12} \ls$ are 
indicated by asterisks ($*$).}

\enddata
\end{deluxetable}

\clearpage
 
\figcaption{ (a) The distribution of infrared luminosities in our HCN 
survey sample of 53 galaxies plus a dozen galaxies with published
HCN observations. More than half of the sample 
galaxies have $L_{\rm IR} < 10^{11} \ls$ and most of them 
have $L_{\rm IR} < 3\times 10^{10} \ls$. 
(b) The distribution of CO luminosities in the same 65 galaxies.
(c) The distribution of HCN luminosities in the same sample. 
\label{fig1}}

\figcaption{ The HCN spectra observed with the IRAM 30m telescope. 
The vertical axis is
the main-beam temperature scale in units of mK. For nearby galaxies, 
only the central spectra are shown here. CO spectra
(dashed lines) from the same telescope have been divided by indicated 
factors for comparison. The spectra
are ordered by right ascension, as in Table~1. 
Some spectra are CO J=2-1, which have a much
smaller beam size than the HCN J=1-0 observations. Some frames 
have only the CO line emission windows (derived from published 
data) indicated. For NGC 253, we show the CO J=2-1
spectrum obtained from the NRAO 12m (scaled to 
the main-beam temperature scale for proper comparison), which has 
roughly the same beam size as that of the HCN J=1-0 from the IRAM 30m.
\label{fig2}}

\figcaption{ HCN spectra obtained from the NRAO 12m 
telescope. The temperature scale is $T_R^*$ in mK. 
$T_{mb}=T_R^*/\eta_M^*$, with a corrected beam efficiency 
$\eta_M^* \sim 0.89$. As in Fig.~2, only the central spectra 
are shown here for nearby galaxies and a few of them (NGC~253, 
NGC~2903, M82, and NGC~6946) are the observations done at 
both the 12m and 30m telescopes. The dashed lines are CO J=1-0 
spectra divided 
by indicated factors. The ultraluminous infrared galaxy 
IRAS 17208$-$0014 (the only one observed with the 12m) is 
shown last. The spectra of NGC~1068 (obtained
from the FCRAO 14m) and NGC~1055 (not detected) are not shown.
\label{fig3}}

\figcaption{ HCN and IR luminosities in 
our HCN survey sample ({\it open circles}) compared with the original
correlation proposed by Solomon \etal (1992) for only 10 galaxies
({\it circles with asterisks}). Note that we used the IR luminosity rather than
the far-IR luminosity as in Solomon \etal (1992). Representative 
error bar is shown for the highest \lhcn~ in the sample, which
includes $\sim$20\% calibration uncertainty.
The line indicates $logL_{\rm IR}=0.97logL_{\rm HCN}+3.1$ (see text). 
\label{fig4}} 

\figcaption{ (a) Relationship between the HCN luminosity and the luminosity
ratio of the 12~$\mu$m-to-IR emission. (b) The HCN-to-CO
luminosity ratio is plotted against the ratio of 
the 12-to-100$\mu$m emission. There are no significant 
correlations found implying
that galaxies with excess mid-IR emission do not show stronger
HCN emission and do not contain higher fractions of dense molecular gas.
The sample is divided into luminous and ultraluminous galaxies
(LIGs/ULIGs) with $L_{\rm IR} \ge 10^{11} \ls$ 
({\it filled circles}) and less luminous galaxies ({\it open circles}). 
Representative error bar is also shown for the highest \lhcn.
\label{fig5}}

\clearpage

\begin{figure*}
\plotone{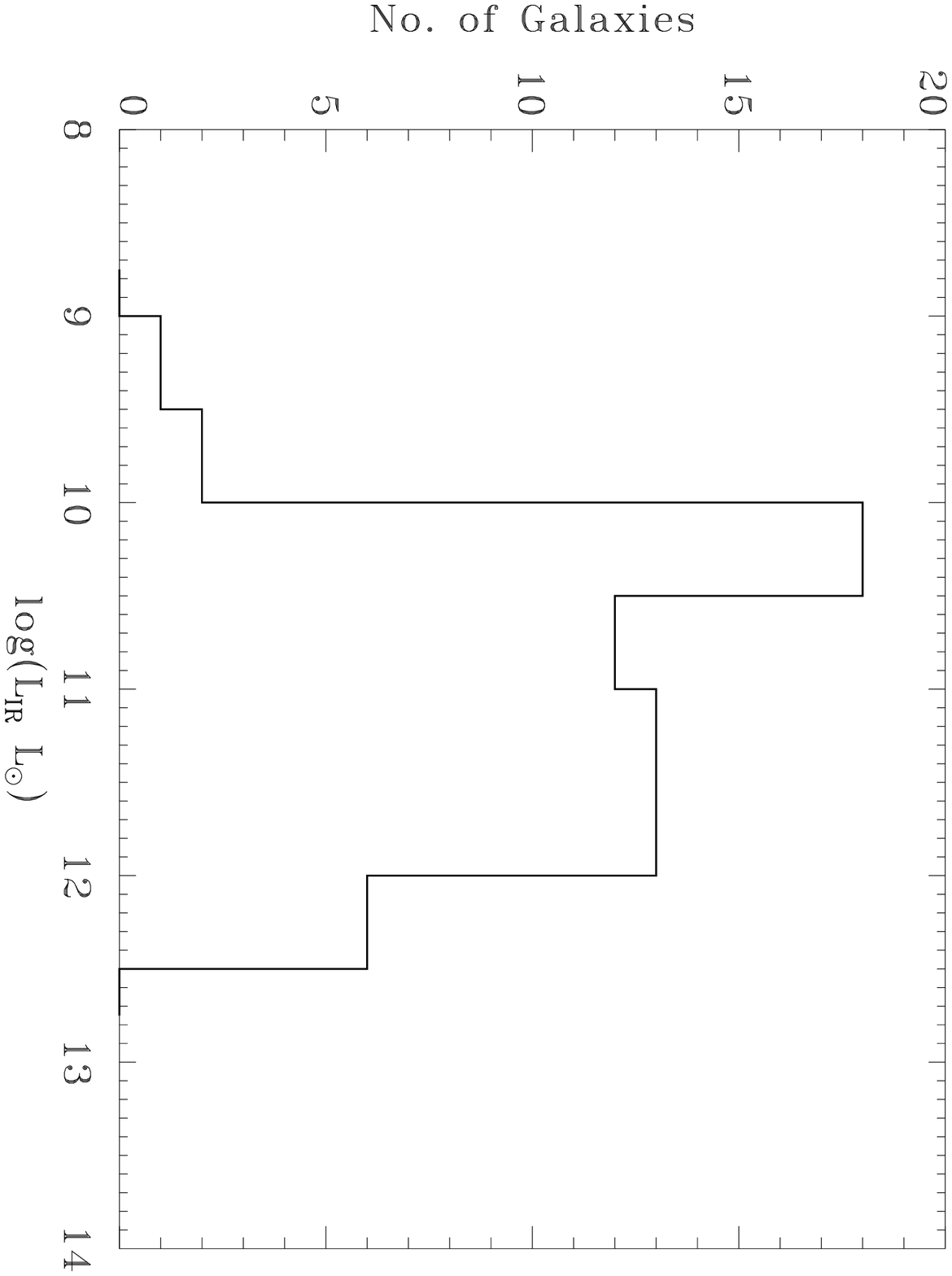}
Fig.~1a
\end{figure*}
\begin{figure*}
\plotone{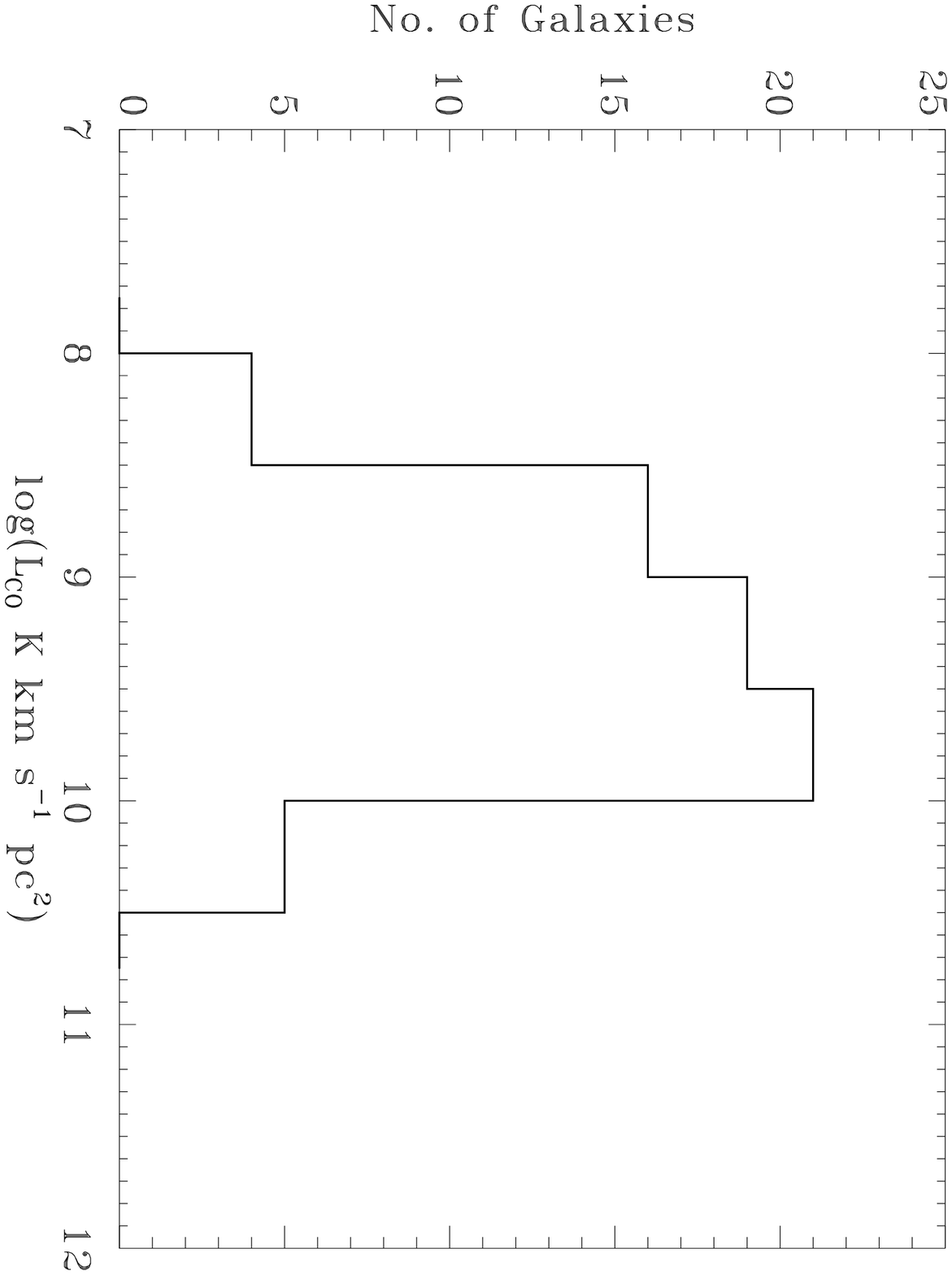}
Fig.~1b
\end{figure*}
\begin{figure}
\plotone{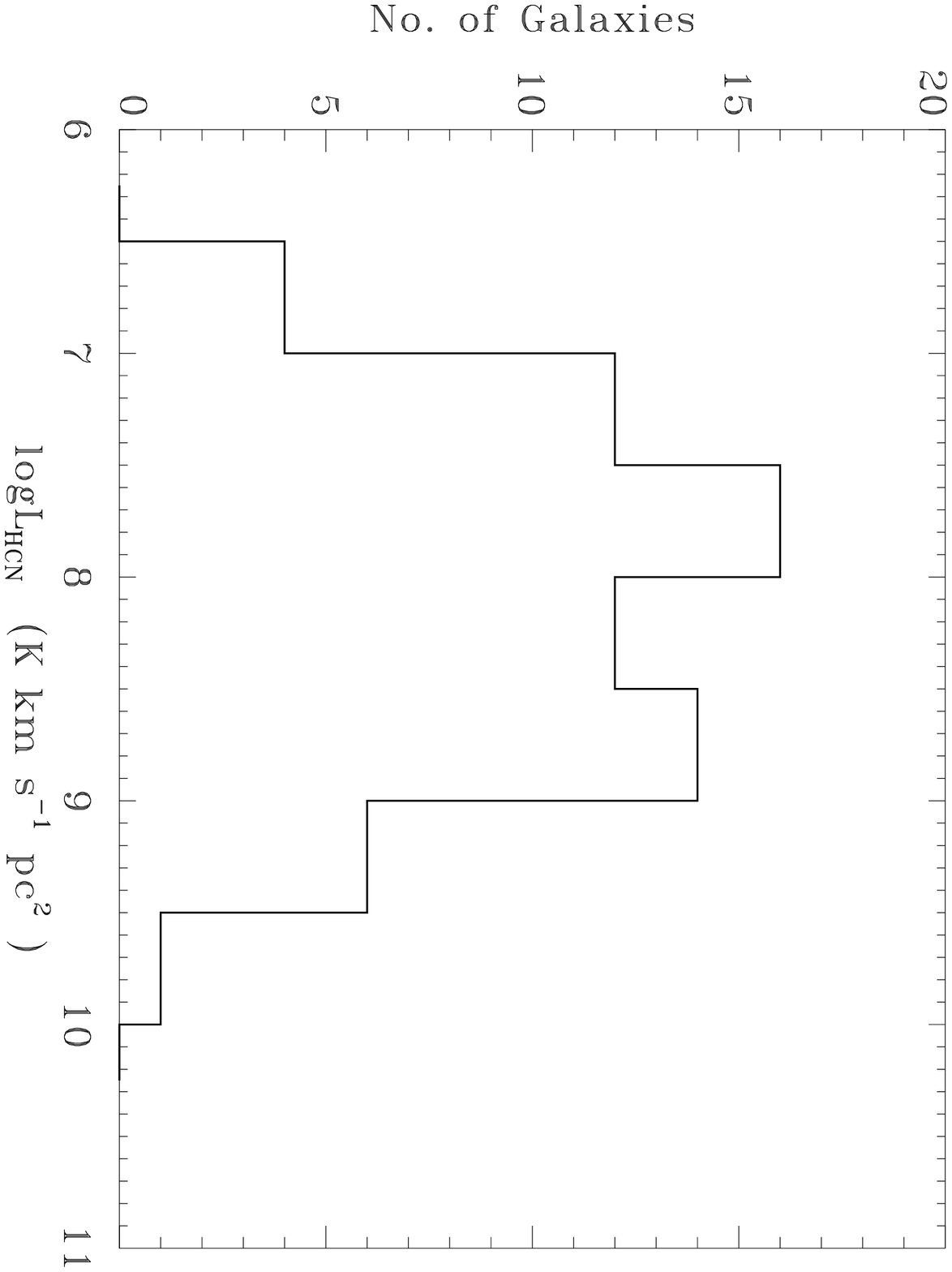}
Fig.~1c
\end{figure}

\begin{figure}
Fig.~2a
See  0310341.fg2a.gif

\vskip 1in

Fig.~2b

See  0310341.fg2b.gif
\vskip 1in

Fig.~2c

See  0310341.fg2c.gif

\end{figure}

\begin{figure}
Fig.~3a

See   0310341.fg3a.gif

\vskip 1in

Fig.~3b

See   0310341.fg3b.gif
\end{figure}
\begin{figure}
\epsscale{0.7}
\plotone{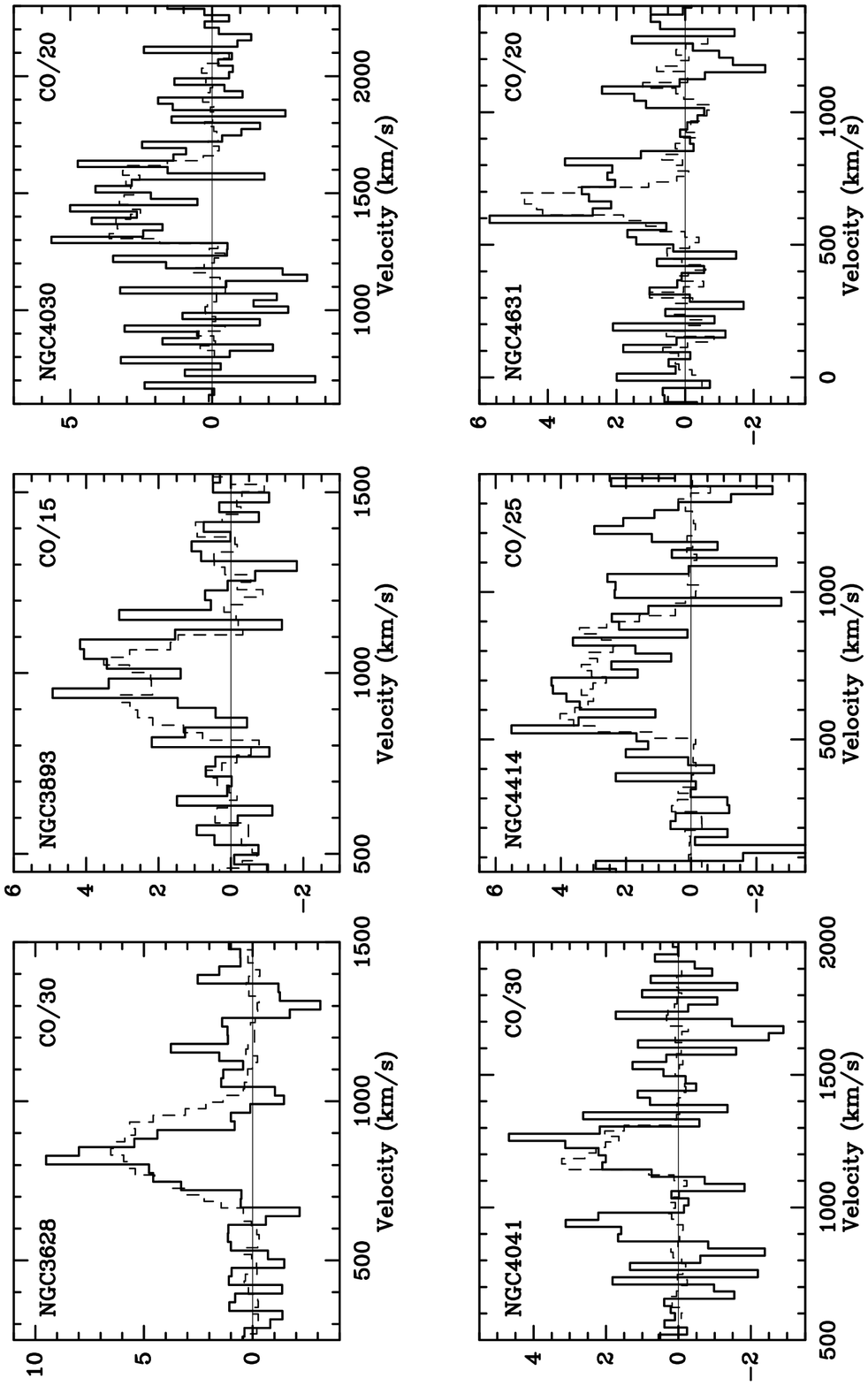}
Fig.~3c
\end{figure}
\begin{figure}
\epsscale{0.7}
\plotone{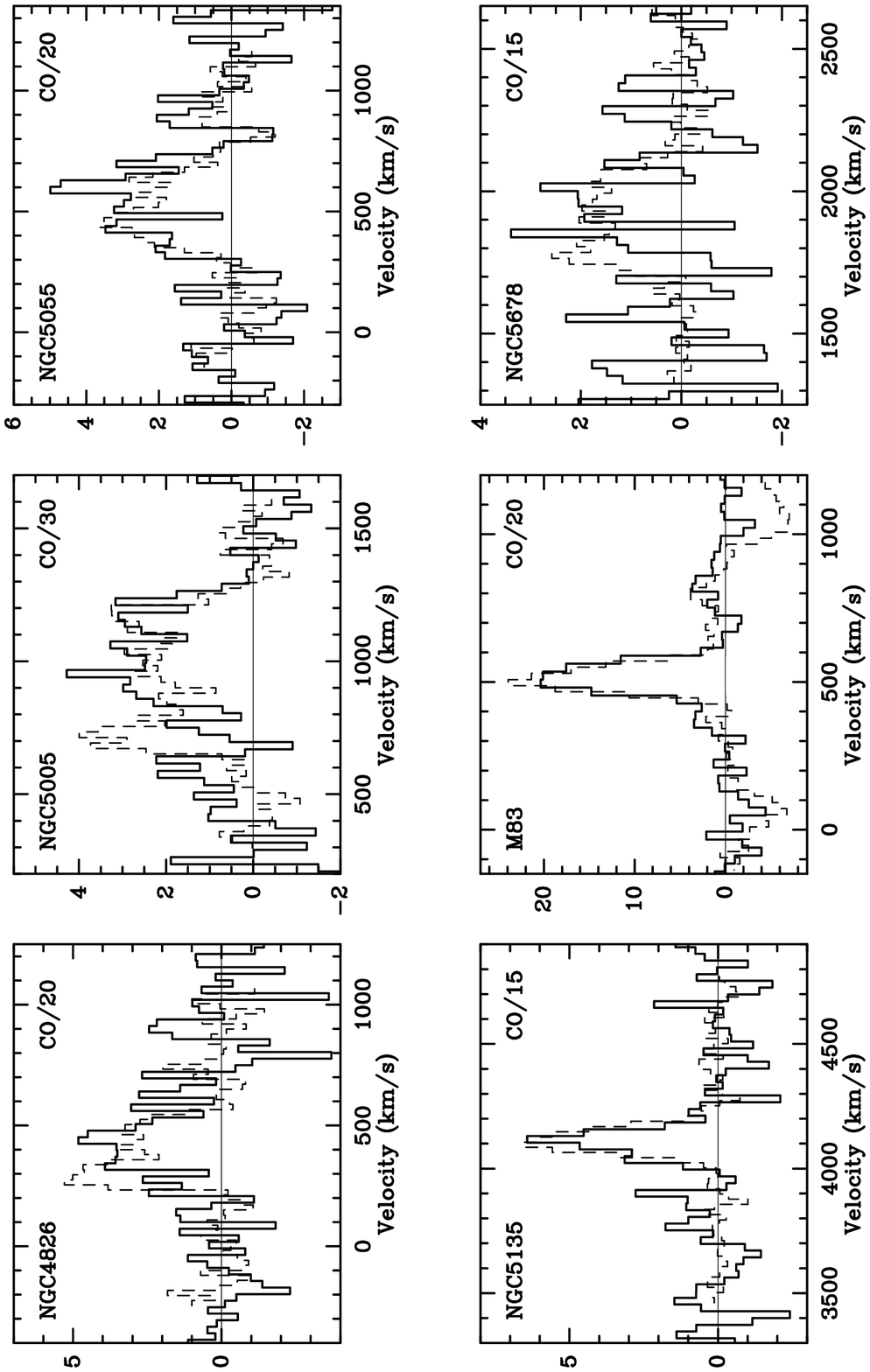}
Fig.~3d
\end{figure}
\begin{figure}
\epsscale{0.7}
\plotone{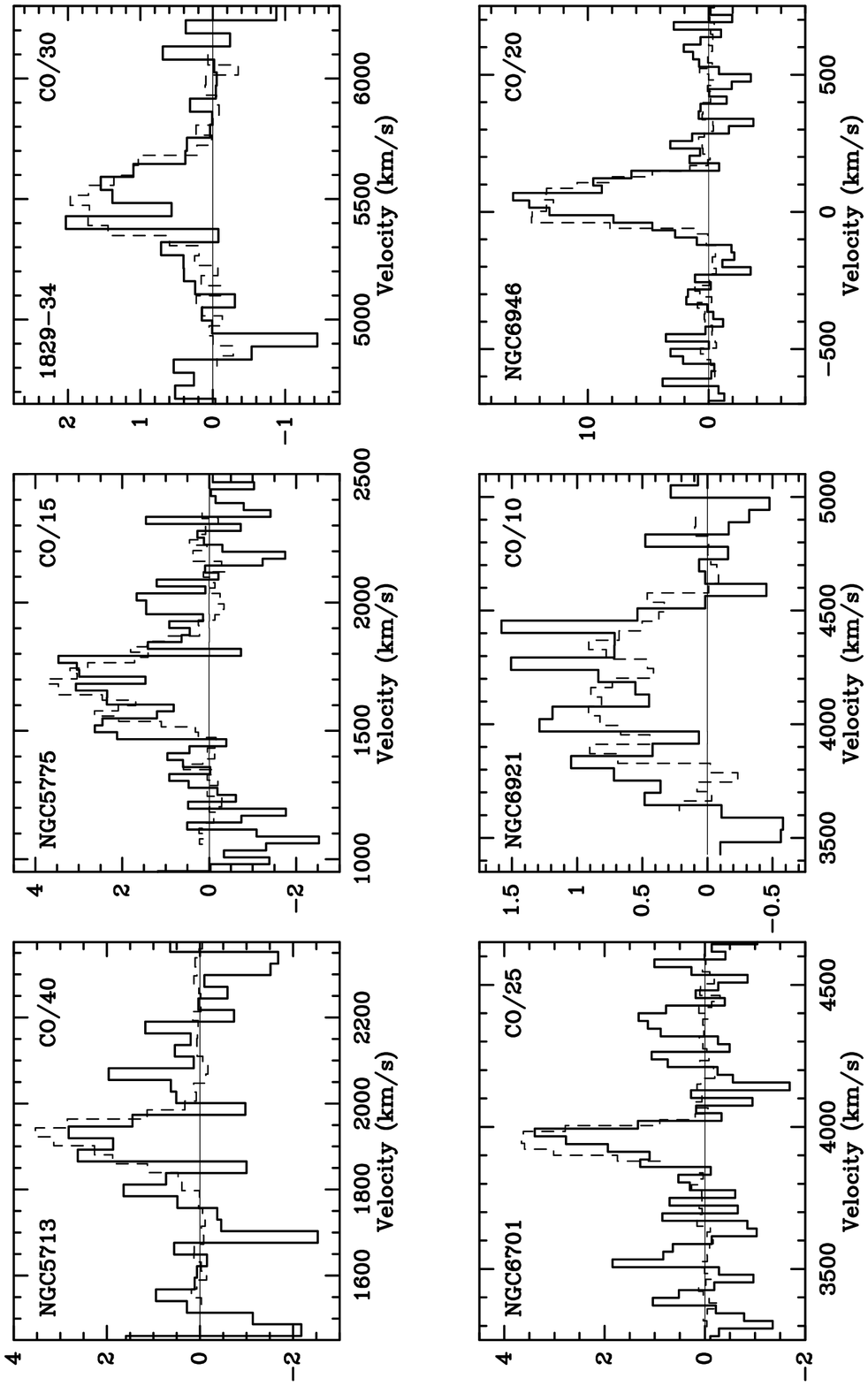}
Fig.~3e
\end{figure}
\begin{figure}
\epsscale{0.7}
\plotone{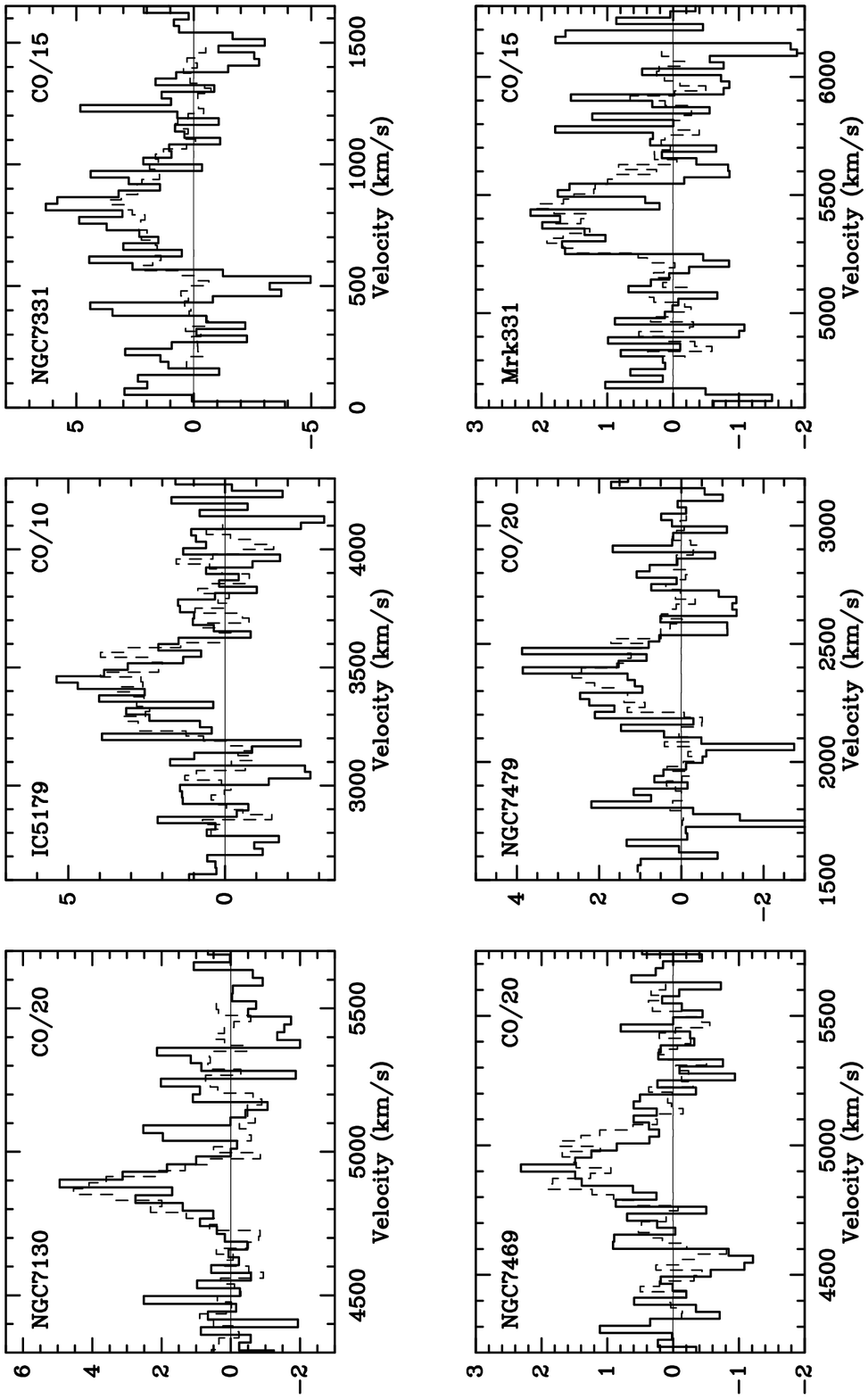}
Fig.~3f
\end{figure}
\begin{figure}
\epsscale{1}
\plotone{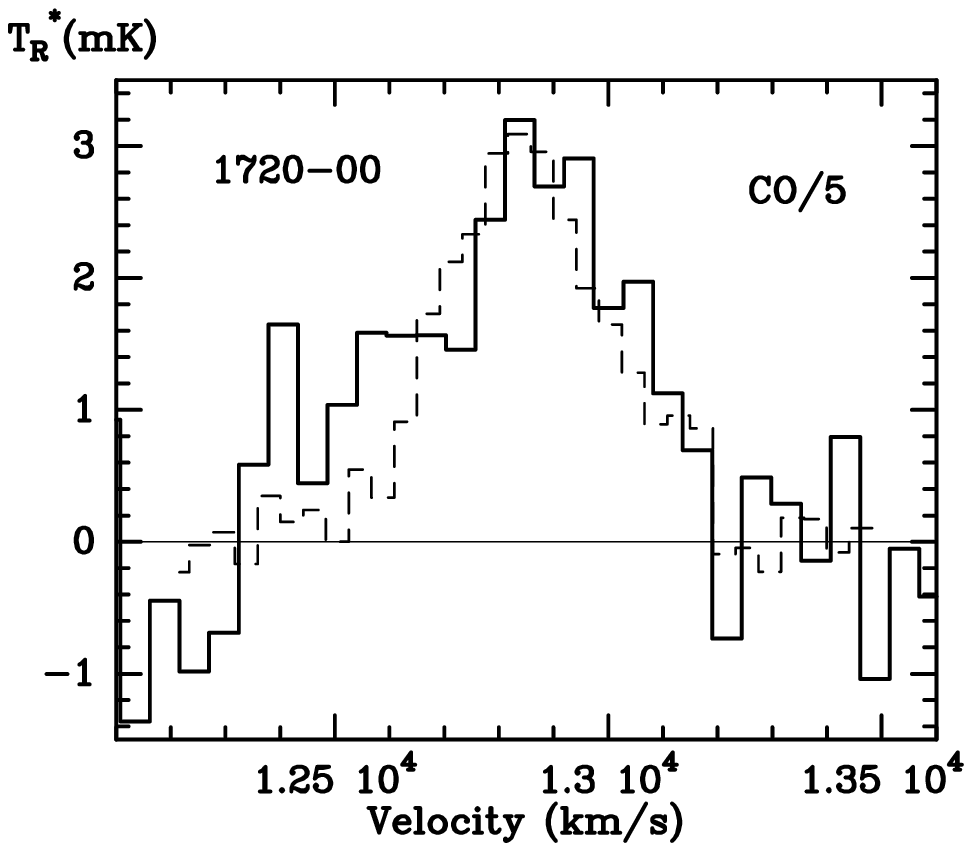}
Fig.~3g
\end{figure}

\begin{figure}
\plotone{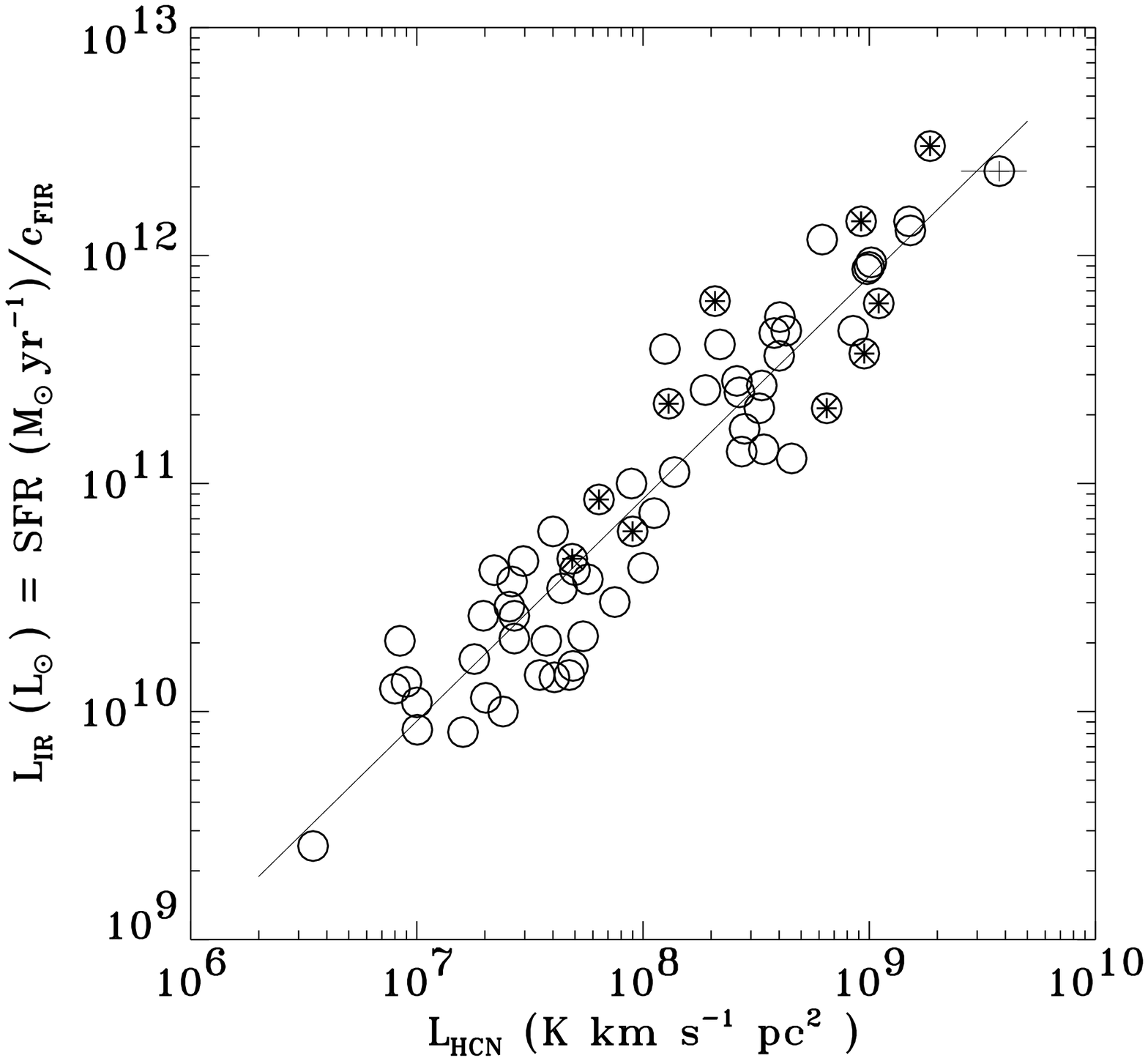}
Fig.~4
\end{figure}

\begin{figure}
\plotone{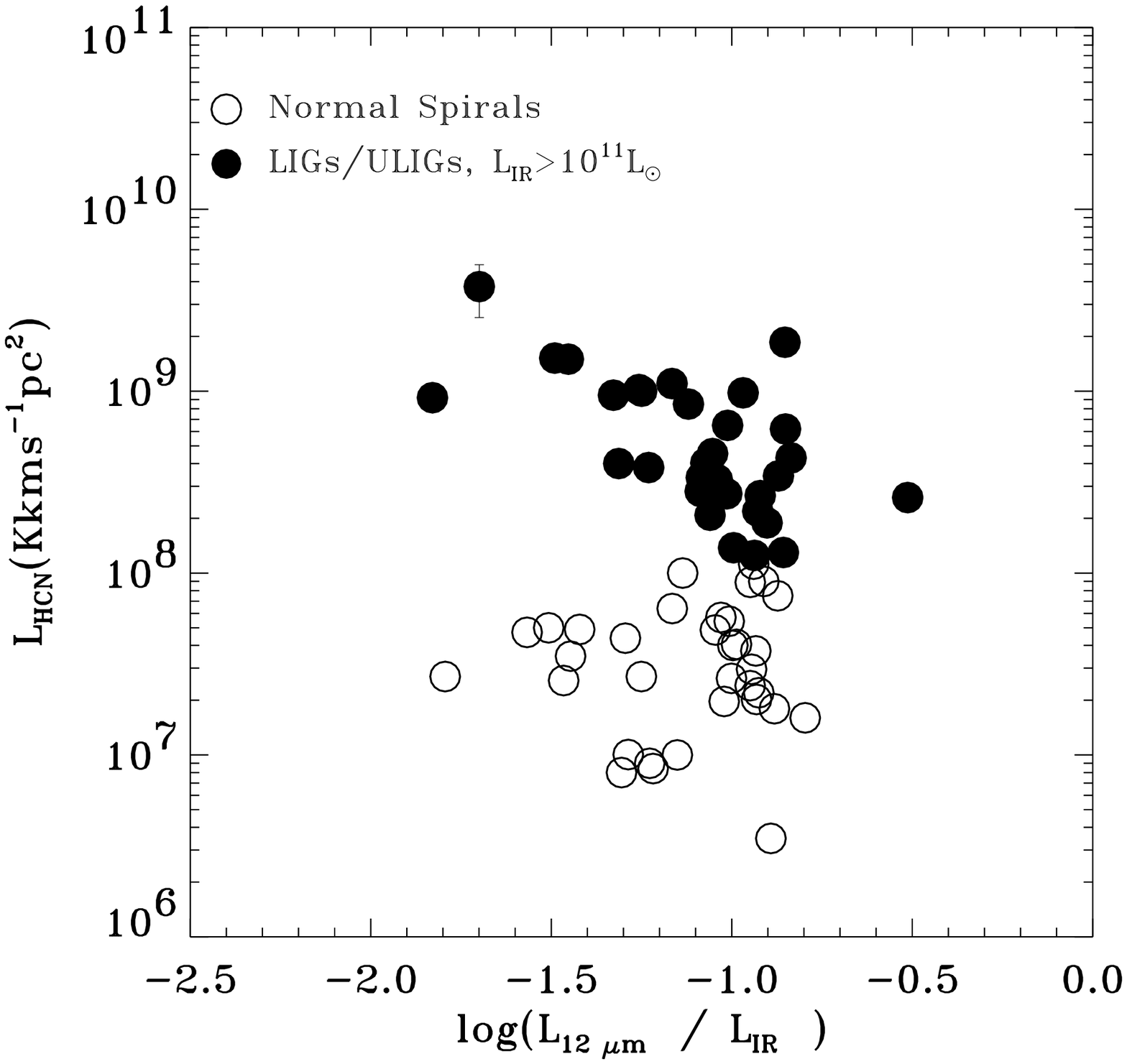}
Fig.~5a
\end{figure}
\begin{figure}
\plotone{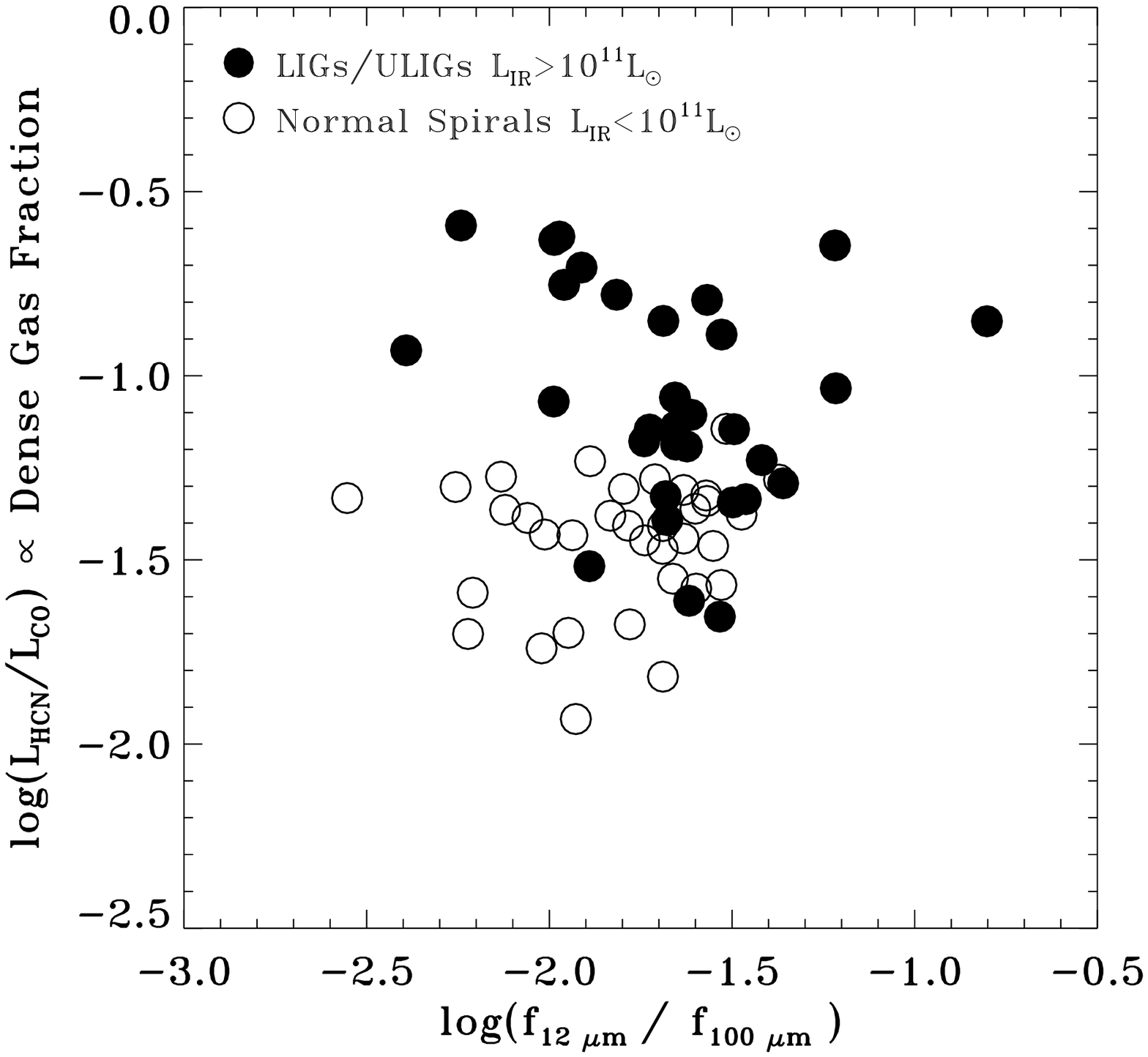}
Fig.~5b
\end{figure}

\end{document}